\documentstyle[pra,aps,psfig]{revtex}
\begin{document}
\draft
\tighten

\title{Quantum Maxwell-Bloch equations\\
       for spatially inhomogeneous semiconductor lasers}

\author{Holger F. Hofmann and Ortwin Hess}
\address{Theoretical Quantum Electronics,Institute of Technical Physics, DLR\\
Pfaffenwaldring 38--40, D--70569 Stuttgart, Germany}

\date{\today}

\maketitle

\begin{abstract}
We present quantum Maxwell-Bloch equations (QMBE) for spatially
inhomogeneous
semiconductor laser devices. The QMBE are derived from fully
quantum mechanical operator dynamics describing the interaction of
the light field with the quantum states of the electrons and the holes
near the band gap. By taking into account field-field correlations
and field-dipole correlations, the QMBE include quantum noise
effects which cause spontaneous emission and amplified spontaneous
emission. In particular, the source of spontaneous emission is obtained by
factorizing the dipole-dipole correlations into a product of
electron and hole densities.
The QMBE are formulated for general devices,
for edge emitting lasers and for vertical cavity surface emitting lasers,
providing a starting point for the detailed analysis of spatial coherence
in the near field and far field patterns of such laser diodes.
Analytical expressions are given for the spectra of gain and
spontaneous emission described by the QMBE. These results are applied to
the case of a broad area laser, for which the frequency and
carrier density dependent spontaneous emission factor $\beta$ and the
evolution of the far field pattern near threshold are derived.
\end{abstract}


\section{introduction}

The spatiotemporal dynamics of semiconductor lasers can be simulated
successfully by semiclassical Maxwell-Bloch equations
without including any quantum effects in the light field
(for an overview of the theory and modeling see \cite{OH96pqe} and references therein).
The classical treatment of the light field is justified by the high intensity of
the laser light well above threshold. However, the incoherent noise
required by the uncertainty principle in both the electrical dipole of the
semiconductor medium and the light field itself is of significant importance
when several cavity modes compete or when the laser is close to threshold.

Photon rate equations for multi-mode operation of semiconductor lasers
show that the spontaneous emission terms may contribute significantly to the
spectral characteristics of the light field emitted by the laser
\cite{Ebe93,Lee82}. Such models assume a fixed mode structure determined
entirely by the empty cavity. This assumption does not apply to
gain guided lasers and to unstable resonators, however
\cite{Pet79,Yar82,Sie96}. In these cases it is therefore desirable to
explicitly describe the spatial coherence of spontaneous emission.

The spatial coherence of spontaneous emission and amplified spontaneous
emission is even more important in devices close to threshold or devices
with a light field output dominated by spontaneous emission such as
superluminescent diodes and ultra low threshold semiconductor lasers
\cite{Gun94,Yam93}. Ultra-low threshold lasers may actually operate in a
regime of negative gain where spontaneous emission is the only source of
radiation \cite{Gun94}. A description of the light field emitted by such
devices therefore requires an explicit description of the spatial
coherence in spontaneous emission as well.

An approach to the consistent inclusion of the quantum noise properties
of the light field in the dynamics of semiconductor laser diodes
using nonequilibrium Green's functions has been presented in
\cite{Jah93,Jah95,Hen96}.
In this approach, the linear optical response of the medium is varied
as a function of the time-dependent electron-hole distributions.
Although the non-equilibrium Green's function presents an elegant solution
for the description of many-body effects\cite{Jah93}, the
representation of the interband dipole dynamics by Green's functions
causes a non-Markovian memory effect which is difficult to handle and
is therefore usually neglected \cite{Jah95}. Moreover, the need to
determine the Green's function corresponding to the dynamically varying
carrier distribution
requires a computational effort far greater than that required for the
integration of the corresponding Maxwell-Bloch equations. Therefore, as
stated in \cite{Hen96}, an exact analytical investigation of the spatial
mode structures in realistic cavities using non-equilibrium Green's functions
is out of reach.
In order to simulate the spatiotemporal dynamics of multi mode operation,
of lasers near threshold,
low threshold lasers or superluminescent diodes, it is therefore desireable
to formulate an alternative approach to the problem of spontaneous
emission and amplified spontaneous emission in such devices which is based on
Maxwell-Bloch equations.
By including the spatiotemporal dynamics of the
interband dipole in such equations, non-Markovian terms are avoided and
the quantum mechanical equations may be integrated in a straightforward manner.

The starting point for our description of quantum noise effects is the
dynamics of quantum mechanical operators of the
field and carrier system. Since the operator dynamics of the
carrier system have been investigated in the context of Maxwell-Bloch
equations before \cite{Hes96} and the light field equations correspond
exactly to the classical Maxwell's equations, it is possible
to focus only on the local light-matter interaction. Once the properties
of this interaction are formulated in terms of the expectation values of
field-field correlations, dipole-field correlations, carrier densities,
fields and dipoles, the dynamics of the carrier system and the light field
propagation may be added.

In section \ref{sec:lci}, the quantum dynamics of the interaction between
the light field and the carrier system is formulated in terms of Wigner
distributions for the carriers and of spatially continuous amplitudes
for the light field. The equations are formulated for both bulk material
and for quantum wells including the effects of anisotropic coupling to
the polarization components of the light field.
Section \ref{sec:cdyn} summarizes the effects of the dynamics of the
electron-hole system in the semiconductor material.
The light field dynamics are introduced in section \ref{sec:Max}.
By quantizing Maxwell's equation, the coupling constant $g_0$
introduced in section \ref{sec:lci} is expressed in terms of the
interband dipole matrix element.
The complete set of quantum Maxwell-Bloch equations is presented in
section \ref{sec:QMBE}. Based on this general formulation, specific
approximate versions for quantum well edge emitting and vertical cavity
surface emitting lasers are derived.
The possibility of including two time correlations in the quantum
Maxwell-Bloch equations is discussed and equations are given for the
case of vertical cavity surface emitting lasers.
In section \ref{sec:ase}, analytical
results for the spectra of gain and spontaneous emission in quantum wells
as well as the spontaneous emission factor $\beta$
and the far field pattern of amplified spontaneous emission
in broad area quantum well lasers are presented.
Section \ref{sec:concl} concludes the article.


\section{Dynamics of the light-carrier interaction}
\label{sec:lci}

\subsection{Hamiltonian dynamics of densities and fields}

In following, we will describe the active semiconductor medium in terms of an isotropic two-band model
where, for the case of the holes, a suitably averaged effective mass
is taken \cite{Hes96}. A generalization to more bands is straightforward.
In terms of the local annihilation operators for photons ($\hat{b}_{\bf R}$),
electrons ($\hat{c}_{\bf R}$), and holes ($\hat{d}_{\bf R}$), the Hamiltonian
of the light-carrier interaction can be written as
\begin{equation}
\label{eq:hamilton}
\hat{H}^{cL}= \hbar g_0 \sum_{\bf R}
\left( \hat{b}^\dagger_{\bf R}\hat{c}_{\bf R}\hat{d}_{\bf R}
+ \hat{b}_{\bf R}\hat{c}^\dagger_{\bf R}\hat{d}^\dagger_{\bf R} \right).
\end{equation}
The operator dynamics associated with this Hamiltonian are then given by
\begin{mathletters}
\begin{eqnarray}
\label{eq:dyn}
\left.\frac{\partial}{\partial t}\hat{b}_{\bf R}\right|_{cL} &=&
         -ig_0\hat{c}_{\bf R}\hat{d}_{\bf R}\\
\left.\frac{\partial}{\partial t}\hat{c}_{\bf R}\hat{d}_{\bf R^\prime} \right |_{cL} &=&
          ig_0 \left(\hat{b}_{\bf R}
                     \hat{d}^\dagger_{\bf R}\hat{d}_{\bf R^\prime} +
                     \hat{b}_{\bf R^\prime}
                     \hat{c}^\dagger_{\bf R^\prime}\hat{c}_{\bf R} -
                     \hat{b}_{\bf R} \delta_{\bf R,R^\prime}\right) \\
\left.\frac{\partial}{\partial t}\hat{c}^\dagger_{\bf R}\hat{c}_{\bf R^\prime} \right |_{cL} &=&
          -ig_0 \left(\hat{b}_{\bf R^\prime}
                      \hat{c}^\dagger_{\bf R}\hat{d}^\dagger_{\bf R^\prime}
                     -\hat{b}^\dagger_{\bf R}
                      \hat{c}_{\bf R^\prime}\hat{d}_{\bf R}\right) \\
\left.\frac{\partial}{\partial t}\hat{d}^\dagger_{\bf R}\hat{d}_{\bf R^\prime} \right |_{cL} &=&
          -ig_0 \left(\hat{b}_{\bf R^\prime}
                      \hat{c}^\dagger_{\bf R^\prime}\hat{d}^\dagger_{\bf R}
                     -\hat{b}^\dagger_{\bf R}
                      \hat{c}_{\bf R}\hat{d}_{\bf R^\prime}\right).
\end{eqnarray}
\end{mathletters}
The discrete positions ${\bf R}$ correspond to the lattice sites of the
Bravais lattice describing the semiconductor crystal. Each lattice site
actually represents the spatial volume $\nu_0$ of
the Wigner-Seitz cell of the lattice. For zincblende crystal structure, this
volume is equal to one quarter of the cubed lattice constant. In the case
of AlGaAs structures the lattice constant is about $5.65 \times 10^{-10}$m
and $\nu_0 \approx 4.5 \times 10^{-29} \mbox{m}^3$ \cite{Yu96}. The
photon annihilation operator $\hat{b}_{\bf R}$ therefore describes the
annihilation of a photon within a volume $\nu_0$.  Here, we will focus on the
light-carrier interaction and the quantum noise contributions responsible
for spontaneous emission. For that purpose, we extend the semiclassical
description by including not only the expectation values of the field and
dipole operators, $\langle \hat{b}_{\bf R}\rangle$ and
$\langle \hat{c}_{\bf R}\hat{d}_{\bf R^\prime}\rangle$, respectively,
but also the field-field correlations
$\langle \hat{b^\dagger}_{\bf R}\hat{b}_{\bf R^\prime}\rangle$ and the
field-dipole correlation $\langle \hat{b}^\dagger_{\bf R}
\hat{c}_{\bf R^\prime}\hat{d}_{\bf R^{\prime\prime}}\rangle$.
The factorized equations of motion then read
\begin{mathletters}
\begin{eqnarray}
\label{eq:factdyn}
\left.\frac{\partial}{\partial t}
\langle \hat{b}^\dagger_{\bf R}\hat{b}_{\bf R^\prime}\rangle  \right |_{cL} &=&
  -ig_0\left(\
  \langle\hat{b}^\dagger_{\bf R}
         \hat{c}_{\bf R^\prime}\hat{d}_{\bf R^\prime}\rangle
 -\langle\hat{b}^\dagger_{\bf R^\prime}
         \hat{c}_{\bf R}\hat{d}_{\bf R}\rangle^*
       \right)
\label{eq:factdynbb}
\\
\left.\frac{\partial}{\partial t}
\langle\hat{b}^\dagger_{\bf R}
       \hat{c}_{\bf R^\prime}\hat{d}_{\bf R^{\prime\prime}}\rangle \right |_{cL} &=&
   ig_0 \left(
  \langle\hat{b}^\dagger_{\bf R}\hat{b}_{\bf R^\prime}\rangle
  \langle\hat{d}^\dagger_{\bf R^\prime}\hat{d}_{\bf R^{\prime\prime}}\rangle
 +\langle\hat{b}^\dagger_{\bf R}\hat{b}_{\bf R^{\prime\prime}}\rangle
  \langle\hat{c}^\dagger_{\bf R^{\prime\prime}}\hat{c}_{\bf R^\prime}\rangle
 -\langle\hat{b}^\dagger_{\bf R}\hat{b}_{\bf R^\prime}\rangle
  \delta_{\bf R^\prime,R^{\prime\prime}}
        \right) \nonumber \\
&& +ig_0 \langle\hat{c}^\dagger_{\bf R}\hat{c}_{\bf R^\prime}\rangle
         \langle\hat{d}^\dagger_{\bf R}\hat{d}_{\bf R^{\prime\prime}}\rangle
\label{eq:factdynbcd}
\\
\left.\frac{\partial}{\partial t}
\langle\hat{c}^\dagger_{\bf R}\hat{c}_{\bf R^\prime}\rangle \right |_{cL} &=&
   ig_0 \left(
  \langle\hat{b}^\dagger_{\bf R}\hat{c}_{\bf R^\prime}\hat{d}_{\bf R}\rangle
 -\langle\hat{b}^\dagger_{\bf R^\prime}
         \hat{c}_{\bf R}\hat{d}_{\bf R^\prime}\rangle^*
        \right)
\label{eq:factdyncc}
\\
\left.\frac{\partial}{\partial t}
\langle\hat{d}^\dagger_{\bf R}\hat{d}_{\bf R^\prime}\rangle \right |_{cL} &=&
   ig_0 \left(
  \langle\hat{b}^\dagger_{\bf R}\hat{c}_{\bf R}\hat{d}_{\bf R^\prime}\rangle
 -\langle\hat{b}^\dagger_{\bf R^\prime}
         \hat{c}_{\bf R^\prime}\hat{d}_{\bf R}\rangle^*
        \right).
\label{eq:factdyndd}
\end{eqnarray}
Note that this set of equations already represents a closed description
of the field dynamics.
If, as in many experimental configurations, the absolute phase of the light field and dipole
operators may be considered unknown, these equations are sufficient for
a description of the light-carrier interaction.
However, when two time correlations are of interest or in the case of coherent excitation by injection of
an external laser it may also be necessary to additionally consider the dynamics of the field and dipole expectation
values, i.e.,
\begin{eqnarray}
\left.\frac{\partial}{\partial t}
\langle\hat{b}_{\bf R}\rangle \right |_{cL} &=&
  -ig_0\langle\hat{c}_{\bf R}\hat{d}_{\bf R}\rangle
\label{eq:factdynb}
\\
\left.\frac{\partial}{\partial t}
\langle\hat{c}_{\bf R}\hat{d}_{\bf R^\prime}\rangle \right |_{cL} &=&
    ig_0 \left(
   \langle\hat{b}_{\bf R}\rangle
   \langle\hat{d}^\dagger_{\bf R}\hat{d}_{\bf R^\prime}\rangle
+  \langle\hat{b}_{\bf R^\prime}\rangle
   \langle\hat{c}^\dagger_{\bf R^\prime}\hat{c}_{\bf R}\rangle
-  \langle\hat{b}_{\bf R}\rangle
   \delta_{\bf R,R^\prime}
         \right).
\label{eq:factdyncd}
\end{eqnarray}
\end{mathletters}

\subsection{Physical background of the factorization}

In the following, we will briefly discuss the implications of the factorization performed in the derivation
of (\ref{eq:factdyn}).
In order to formulate the dynamics of the light-matter interaction
without including higher order correlations, three terms have been factorized
in the time derivative of the field-dipole correlation (\ref{eq:factdynbcd}).
The factorizations are
\begin{mathletters}
\begin{eqnarray}
\label{eq:fact1}
\langle\hat{b}^\dagger_{\bf R}\hat{b}_{\bf R^{\prime\prime}}
       \hat{c}^\dagger_{\bf R^{\prime\prime}}\hat{c}_{\bf R^\prime}\rangle
&\approx&
\langle\hat{b}^\dagger_{\bf R}\hat{b}_{\bf R^{\prime\prime}}\rangle
\langle\hat{c}^\dagger_{\bf R^{\prime\prime}}\hat{c}_{\bf R^\prime}\rangle
\\ \label{eq:fact2}
\langle\hat{b}^\dagger_{\bf R}\hat{b}_{\bf R^\prime}
       \hat{d}^\dagger_{\bf R^\prime}\hat{d}_{\bf R^{\prime\prime}}\rangle
&\approx&
\langle\hat{b}^\dagger_{\bf R}\hat{b}_{\bf R^\prime}\rangle
\langle\hat{d}^\dagger_{\bf R^\prime}\hat{d}_{\bf R^{\prime\prime}}\rangle
\\ \label{eq:fact3}
\langle\hat{c}^\dagger_{\bf R}\hat{c}_{\bf R^\prime}
       \hat{d}^\dagger_{\bf R}\hat{d}_{\bf R^{\prime\prime}}\rangle
&\approx&
\langle\hat{c}^\dagger_{\bf R}\hat{c}_{\bf R^\prime}\rangle
\langle\hat{d}^\dagger_{\bf R}\hat{d}_{\bf R^{\prime\prime}}\rangle.
\end{eqnarray}
\end{mathletters}
No additional factorizations are necessary in the density dynamics of
photons, electrons and holes. The three factorizations are based on the
assumption of statistical
independence between the densities of photons, electrons and holes.
Since we are not considering fluctuations in the particle densities,
this is a necessary assumption.

Equations (\ref{eq:fact1}) and (\ref{eq:fact2}) separate the photon
density from the carrier densities. These terms represent stimulated
emission processes. Therefore, a correlation of the photon density with the carrier
densities would lead to a modified stimulated emission rate. Below
threshold, this effect will be small because the amplified spontaneous
emission is distributed over many modes such that the local correlations
between photon and carrier densities are weak. Above threshold, the
photon number fluctuations in the lasing mode cause relaxation oscillations.
The photon number fluctuations are nearly
ninety degrees out of phase with the carrier number fluctuations.
Therefore, the time averaged correlation is still negligible.

Equation (\ref{eq:fact3}) separates the electron and hole densities.
This term represents the spontaneous emission caused by the simultaneous
presence of electrons and holes in the same location.
Although it is reasonable to assume that the high rate of scattering at
high carrier densities effectively reduce all electron-hole correlations
to zero, it is important to note that the interband dipole $\langle \hat{c}
_{\bf R}\hat{d}_{\bf R^\prime}\rangle$ implies a phase correlation
between the electrons and the holes. In fact, the spontaneous emission
term factorized according to equation (\ref{eq:fact3}) originates from
the dipole-dipole correlation
$\langle \hat{c}^\dagger_{\bf R}\hat{d}^\dagger_{\bf R^\prime}
\hat{c}_{\bf R^{\prime\prime}}\hat{d}_{\bf R^{\prime\prime\prime}}\rangle$.
Note that this term could also be factorized into the product
of dipole operators  $\langle \hat{c}_{\bf R}\hat{d}_{\bf R^\prime}\rangle^*
\langle \hat{c}_{\bf R^{\prime\prime}}\hat{d}_{\bf R^{\prime\prime\prime}}
\rangle$. The dynamics of the field-field and the field-dipole correlations
are then identical to the dynamics of the products of the fields and dipoles.
Therefore, that factorization corresponds to the approximations of the
conventional Maxwell-Bloch equations such as described in \cite{Hes96}
which do not rigorously include spontaneous emission.

Generally, spontaneous emission must always arise from random phase fluctuations.
These are given by the product of electron and hole densities.
While the phase dependent dipole relaxes quickly due to scattering
events, the carrier densities are preserved during scattering. Therefore
\begin{mathletters}
\begin{equation}
\langle \hat{c}_{\bf R}\hat{d}_{\bf R^\prime}\rangle^*
\langle \hat{c}_{\bf R^{\prime\prime}}\hat{d}_{\bf R^{\prime\prime\prime}}
\rangle \ll
\langle \hat{c}^\dagger_{\bf R}\hat{c}_{\bf R^{\prime\prime}}\rangle \langle
\hat{d}^\dagger_{\bf R^\prime}\hat{d}_{\bf R^{\prime\prime\prime}}\rangle
\end{equation}
is usually a good assumption in semiconductor systems. Note that this
assumption does fail in the case of low carrier
densities and high dipole inducing fields. However, this case only occurs
if the light field is injected from an external source. In
semiconductor lasers and in light emitting diodes, the major contribution
to the dipole-dipole correlations stems from the product of
electron and hole densities.
To check the statistical independence of photon, electron, and hole densities,
it is also convenient to check the corresponding inequality for the three
particle coherence represented by the field-dipole correlation,
\begin{equation}
\langle \hat{b}^\dagger_{\bf R}
        \hat{c}_{\bf R^\prime}
        \hat{d}_{\bf R^{\prime\prime}}\rangle^*
\langle \hat{b}^\dagger_{R^{\prime\prime\prime}}
        \hat{c}_{\bf R^{\prime\prime\prime\prime}}
        \hat{d}_{\bf R^{\prime\prime\prime\prime\prime}}\rangle
 \ll
\langle \hat{b}^\dagger_{\bf R}
        \hat{b}_{\bf R^{\prime\prime\prime}}\rangle
\langle \hat{c}^\dagger_{\bf R^\prime}
        \hat{c}_{\bf R^{\prime\prime\prime\prime}}\rangle
\langle \hat{d}^\dagger_{\bf R^{\prime\prime}}
        \hat{d}_{\bf R^{\prime\prime\prime\prime\prime}}\rangle.
\end{equation}
\end{mathletters}
Thus, if a calculation does not fulfill this requirement, particle density
correlations additionally have to be taken into account.

\subsection{Wigner function formulation}

In order to connect the light-carrier interaction to the highly dissipative
carrier transport equations, it is practical to transform the carrier
and dipole densities using Wigner transformations \cite{Wig32}.
Replacing the discrete density matrices by continuous ones obtained
by polynomial interpolation will allow e.g.~for numerical purposes an
arbitrary choice of the discretization scales, which generally will be
much larger than a lattice constant. Analytically, it permits an application
of differential operators. Physically, the particle densities are smooth
functions over distances of several lattice constants. A coherence
length shorter than e.g. ten lattice constants would require ${\bf k}$
states with $\mid{\bf k}\mid$ of at least one twentieth of the Brilloin zone
diameter. In typical laser devices, however, the electrons
and holes all accumulate near the fundamental gap at ${\bf k}=0$.
Therefore, it is simply a matter of convenience to define the continuous
densities such that
\begin{mathletters}
\begin{eqnarray}
\rho^e({\bf r}={\bf R},{\bf r^\prime}={\bf R^\prime}) &=&
 \frac{1}{\nu_0}\langle\hat{c}^\dagger_{\bf R}\hat{c}_{\bf R^\prime}\rangle \\
\rho^h({\bf r}={\bf R},{\bf r^\prime}={\bf R^\prime}) &=&
 \frac{1}{\nu_0}\langle\hat{d}^\dagger_{\bf R}\hat{d}_{\bf R^\prime}\rangle \\
\rho^{dipole}({\bf r}={\bf R},{\bf r^\prime}={\bf R^\prime}) &=&
 \frac{1}{\nu_0}\langle\hat{c}_{\bf R}\hat{d}_{\bf R^\prime}\rangle.
\end{eqnarray}
\end{mathletters}
These continuous functions may then be transformed into Wigner functions
by
\begin{mathletters}
\begin{eqnarray}
f^e({\bf r},{\bf k}) &=& \int d^3{\bf r^\prime} e^{-i{\bf kr^\prime}}
\rho^e\left({\bf r}-\frac{\bf r^\prime}{2},{\bf r}+\frac{\bf r^\prime}{2}\right)
\\
f^h({\bf r},{\bf k}) &=& \int d^3{\bf r^\prime} e^{-i{\bf kr^\prime}}
\rho^h\left({\bf r}-\frac{\bf r^\prime}{2},{\bf r}+\frac{\bf r^\prime}{2}\right)
\\
p({\bf r},{\bf k}) &=& \int d^3{\bf r^\prime} e^{-i{\bf kr^\prime}}
\rho^{dipole}\left({\bf r}-\frac{\bf r^\prime}{2},{\bf r}+\frac{\bf r^\prime}{2}\right).
\end{eqnarray}
\end{mathletters}
The normalization of these Wigner functions has been chosen in such a way
that a value of one represents the maximal phase space density possible
for Fermions, that is one particle per state. Since the density of states
in the six-dimensional phase space given by ${\bf r}$ and ${\bf k}$ is
$1/8\pi^3$, a factor of $1/8\pi^3$ will appear whenever actual carrier
densities need to be obtained from the Wigner functions. However, the
normalization in terms of the maximal possible phase space density is
convenient because it represents the probability that a quantum state in
a given region of phase space is occupied. Therefore, the Wigner distribution
corresponding to the thermal equilibrium of a given particle density is
directly given by the Fermi function.

To deal with the light field dynamics in the same manner, the field and
field-field correlation variables must also be defined on a continuous
length scale. In order to obtain photon densities, we define
\begin{mathletters}
\begin{eqnarray}
\cal E({\bf r}={\bf R}) &=&
 \frac{1}{\sqrt{\nu_0}}\langle\hat{b}_{\bf R}\rangle \\
I({\bf r}={\bf R};{\bf r^\prime}={\bf R^\prime}) &=&
 \frac{1}{\nu_0}\langle\hat{b}^\dagger_{\bf R}\hat{b}_{\bf R^\prime}\rangle.
\end{eqnarray}
\end{mathletters}
Finally, the dipole-field correlation must be defined accordingly, such that
\begin{eqnarray}
\Theta^{corr.}({\bf r}={\bf R};{\bf r^\prime}={\bf R^\prime},
               {\bf r^{\prime\prime}}={\bf R^{\prime\prime}}) &=&
\frac{1}{\sqrt{\nu_0^3}}\langle\hat{b}^\dagger_{\bf R}\hat{c}_{\bf R^\prime}
                               \hat{d}_{\bf R^{\prime\prime}}\rangle \\
C({\bf r};{\bf r^\prime},{\bf k}) &=&
        \int d^3{\bf r^{\prime\prime}} e^{-i{\bf kr^{\prime\prime}}}
\Theta^{corr.}\left({\bf r};{\bf r^\prime}-\frac{\bf r^{\prime\prime}}{2},
                       {\bf r^\prime}+\frac{\bf r^{\prime\prime}}{2}      \right).
\end{eqnarray}
With these new definitions, the light-carrier interaction dynamics can
now be expressed in a form which considers both the position and the momentum
of the electrons and holes. The dynamics of emission and absorption
now reads
\begin{mathletters}
\begin{eqnarray}
\left.\frac{\partial}{\partial t} I({\bf r};{\bf r^\prime})\right |_{cL} &=&
-i g_0\frac{\sqrt{\nu_0}}{8\pi^3}\int d^3{\bf k}
\left(C({\bf r};{\bf r^\prime},{\bf k})
     -C^*({\bf r^\prime};{\bf r},{\bf k})\right)
\\
\left.\frac{\partial}{\partial t} C({\bf r};{\bf r^\prime},{\bf k})\right |_{cL}
&=&
ig_0\sqrt{\nu_0}\;\frac{1}{8\pi^3}\!\int\! d^3 {\bf x}\!\int\!d^3 {\bf q} \; e^{i{\bf qx}}
\nonumber \\
&&
\times
\left(f^e\left({\bf r^\prime},{\bf k}+\frac{\bf q}{2}\right)  +
      f^h\left({\bf r^\prime},-{\bf k}+\frac{\bf q}{2}\right) - 1 \right)
I({\bf r};{\bf r^\prime}+{\bf x}) \nonumber
\\
& + &
ig_0\sqrt{\nu_0}\;\frac{1}{8\pi^3}\!\int\! d^3 {\bf q}\;
e^{i{\bf q}({\bf r}-{\bf r^\prime})}
\nonumber \\ &&
\times
f^e\left(\frac{{\bf r}+{\bf r^\prime}}{2},{\bf k}+\frac{\bf q}{2}\right)
f^h\left(\frac{{\bf r}+{\bf r^\prime}}{2},-{\bf k}+\frac{\bf q}{2}\right)
\\
\left.\frac{\partial}{\partial t} f^e({\bf r},{\bf k})\right |_{cL}
&=&
i g_0\sqrt{\nu_0}\;
\frac{1}{8\pi^3}\!\int\! d^3 {\bf x}\!\int\!d^3 {\bf q}\;e^{i{\bf qx}}
\nonumber \\
&& \times
\left( C  \left({\bf r}+{\bf x};{\bf r},{\bf k}+\frac{\bf q}{2}\right)
     - C^*\left({\bf r}+{\bf x};{\bf r},{\bf k}+\frac{\bf q}{2}\right)\right)
\\
\left.\frac{\partial}{\partial t} f^h({\bf r},{\bf k})\right |_{cL}
&=&
i g_0\sqrt{\nu_0}\;
\frac{1}{8\pi^3}\!\int\! d^3 {\bf x}\!\int\!d^3 {\bf q}\;e^{i{\bf qx}}
\nonumber \\
&& \times
\left(C  \left({\bf r}+{\bf x};{\bf r},-{\bf k}+\frac{\bf q}{2}\right)
     -C^*\left({\bf r}+{\bf x};{\bf r},-{\bf k}+\frac{\bf q}{2}\right)\right)
\\[1cm]
\left.\frac{\partial}{\partial t}{\cal E} ({\bf r})\right |_{cL}
&=&
-i g_0\frac{\sqrt{\nu_0}}{8\pi^3}\int d^3{\bf k}\; p({\bf r},{\bf k})
\\
\left.\frac{\partial}{\partial t} p({\bf r},{\bf k})\right |_{cL}
&=&
ig_0\sqrt{\nu_0}\;\frac{1}{8\pi^3}\!\int\! d^3 {\bf x}\!\int\!d^3 {\bf q}\;
e^{i{\bf qx}}
\nonumber \\ && \times
             \left(f^e\left({\bf r},{\bf k}+\frac{\bf q}{2}\right)+
                   f^h\left({\bf r},-{\bf k}+\frac{\bf q}{2}\right)-1\right)
{\cal E}({\bf r}+{\bf x}).
\end{eqnarray}
\end{mathletters}

\subsection{Local approximation}

The integrals over ${\bf x}$ and ${\bf q}$ represent seemingly non-local
effects introduced by the transformation into Wigner functions.
This property of the Wigner transformation retains the coherent effects
in the carrier system. For the interaction of the carriers with the light
field, it ensures momentum conservation by introducing a non-local phase
correlation in the dipole field corresponding to the total momentum of
the electron and hole concentrations involved. Effectively, the integral
over ${\bf q}$ converts the momentum part of the Wigner distributions
into a coherence length. This coherence length then reappears in the
spatial structure of the dipole field and the electromagnetic field
generated by the carrier distribution. However, the coherence length in
the carrier system is usually much shorter than the optical wavelength.
It can therefore be approximated by a spatial delta function. Here, we do this
by noting that
\begin{equation}
\frac{1}{8\pi^3}\!\int\!d^3{\bf q}\;e^{i{\bf qx}} = \delta ({\bf x}).
\end{equation}
If the effects of the momentum shift ${\bf q}$ in the Wigner functions is
neglected, the integrals may then be solved, yielding only local interactions
between the carrier system and the light field:
\begin{mathletters}
\begin{eqnarray}
\label{eq:bulk}
\left.\frac{\partial}{\partial t} I({\bf r};{\bf r^\prime})\right |_{cL} &=&
-i g_0\frac{\sqrt{\nu_0}}{8\pi^3}\int d^3{\bf k}
\left(C({\bf r};{\bf r^\prime},{\bf k})
     -C^*({\bf r^\prime};{\bf r},{\bf k})\right)
\\
\left.\frac{\partial}{\partial t} C({\bf r};{\bf r^\prime},{\bf k})\right |_{cL} &=&
ig_0\sqrt{\nu_0}\;
\left(f^e({\bf r^\prime},{\bf k})+
                   f^h({\bf r^\prime},-{\bf k})-1\right)
I({\bf r};{\bf r^\prime}) \nonumber \\
& + & ig_0\sqrt{\nu_0}\;\delta({\bf r}-{\bf r^\prime})
f^e({\bf r},{\bf k})
f^h({\bf r},-{\bf k})
\\
\left.\frac{\partial}{\partial t} f^e({\bf r},{\bf k})\right |_{cL} &=&
i g_0\sqrt{\nu_0}\;
\left(C({\bf r};{\bf r},{\bf k})
     -C^*({\bf r};{\bf r},{\bf k})\right)
\\
\left.\frac{\partial}{\partial t} f^h({\bf r},{\bf k})\right |_{cL} &=&
i g_0\sqrt{\nu_0}\;
\left(C({\bf r};{\bf r},-{\bf k})
     -C^*({\bf r};{\bf r},-{\bf k})\right)
\\[1cm]
\left.\frac{\partial}{\partial t}{\cal E} ({\bf r})\right |_{cL} &=&
-i g_0\frac{\sqrt{\nu_0}}{8\pi^3}\int d^3{\bf k}\; p({\bf r},{\bf k})
\\
\label{eq:endbulk}
\left.\frac{\partial}{\partial t} p({\bf r},{\bf k})\right |_{cL} &=&
ig_0\sqrt{\nu_0}\;
\left(f^e({\bf r},{\bf k})+f^h({\bf r},-{\bf k})-1\right){\cal E}({\bf r}).
\end{eqnarray}
\end{mathletters}
These equations now provide a compact description of the light-carrier
interaction in a three dimensional semiconductor medium, including the
incoherent quantum noise
term which is the source of spontaneous emission.

\subsection{Light-carrier interaction for quantum wells}

Similar equations may also
be formulated for a quantum well structure by replacing the phase space
density of $1/8\pi^3$ with $1/4\pi^2$, reducing the spatial coordinates
of the carrier system to two dimensions, and introducing a delta function
for the coordinate perpendicular to the quantum well at the points where
field coordinates correspond to dipole coordinates. Of course, the electromagnetic
field remains three dimensional, even though the dipole it originates
from is confined to two dimensions. In particular, the correlation
$C({\bf r};{\bf r^\prime},{\bf k})$ has both a three dimensional
coordinate ${\bf r}$ and a two dimensional coordinate ${\bf r^\prime}$.
It is therefore useful to distinguish the two dimensional and the three
dimensional coordinates. In the following, the two dimensional carrier
coordinates will be marked with the index $\parallel$. Note that in some
cases, both ${\bf r}$ and ${\bf r}_\parallel$ appear in the equations.
In those cases, the in-plane coordinates $r_x$ and $r_y$ are equal, while
the perpendicular coordinate $r_z$ must be equal to the quantum well
coordinate $z_0$.
The equations for the interaction of the three dimensional light field
with the two dimensional electron-hole system in a single quantum well
subband then read
\begin{mathletters}
\begin{eqnarray}
\label{eq:QW}
\left.\frac{\partial}{\partial t} I({\bf r};{\bf r^\prime})\right |_{cL} &=&
-i g_0\frac{\sqrt{\nu_0}}{4\pi^2}\int d^2{\bf k}_\parallel
\left(C({\bf r};{\bf r^\prime}_\parallel,{\bf k}_\parallel)
\delta(r_z^\prime-z_0) \right.
\nonumber \\ & & \left.
     -C^*({\bf r^\prime};{\bf r}_\parallel,{\bf k}_\parallel)
\delta(r_z-z_0)\right)
\\
\left.\frac{\partial}{\partial t} C({\bf r};{\bf r^\prime}_\parallel,
                              {\bf k}_\parallel)\right |_{cL} &=&
 ig_0\sqrt{\nu_0}\;
\left(f^e({\bf r^\prime}_\parallel,{\bf k}_\parallel)+
                   f^h({\bf r^\prime}_\parallel,-{\bf k}_\parallel)-1\right)
I({\bf r};{\bf r^\prime})_{r_z^\prime=z_0} \nonumber \\ &&
+ig_0\sqrt{\nu_0}\;\delta({\bf r}_\parallel-{\bf r^\prime}_\parallel)
\delta(r_z-z_0)
f^e({\bf r}_\parallel,{\bf k}_\parallel)
f^h({\bf r}_\parallel,-{\bf k}_\parallel)
\\
\left.\frac{\partial}{\partial t} f^e({\bf r}_\parallel,{\bf k}_\parallel)\right |_{cL} &=&
 i g_0\sqrt{\nu_0}\;
\left(C({\bf r};{\bf r}_\parallel,{\bf k}_\parallel)_{r_z=z_0}
     -C^*({\bf r};{\bf r}_\parallel,{\bf k}_\parallel)_{r_z=z_0}\right)
\\
\left.\frac{\partial}{\partial t} f^h({\bf r}_\parallel,{\bf k}_\parallel)\right |_{cL} &=&
 i g_0\sqrt{\nu_0}\;
\left(C({\bf r};{\bf r}_\parallel,-{\bf k}_\parallel)_{r_z=z_0}
     -C^*({\bf r};{\bf r}_\parallel,-{\bf k}_\parallel)_{r_z=z_0}\right)
\\[1cm]
\left.\frac{\partial}{\partial t}{\cal E} ({\bf r})\right |_{cL} &=&
-i g_0\frac{\sqrt{\nu_0}}{4\pi^2}\int d^2{\bf k}_\parallel\;
p({\bf r}_\parallel,{\bf k}_\parallel)\delta(r_z-z_0)
\\ \label{eq:endQW}
\left.\frac{\partial}{\partial t} p({\bf r}_\parallel,{\bf k}_\parallel)\right |_{cL} &=&
 ig_0\sqrt{\nu_0}\;
\left(f^e({\bf r}_\parallel,{\bf k}_\parallel)
     +f^h({\bf r}_\parallel,-{\bf k}_\parallel)-1\right)
{\cal E}({\bf r})_{r_z=z_0}.
\end{eqnarray}
\end{mathletters}
Note that the value of $g_0$ will usually be slightly lower than
the bulk value because the overlap of the spatial wavefunctions
of the electrons and the holes in the lowest subbands is less than one.
The equations derived above represent the interaction of a single conduction
band and a single valence band with a single scalar light field.
Neither the spin degeneracy of the carriers nor the polarization of the
light field has been considered.

\subsection{Spin degeneracy and light field polarization}

Since the geometry of light field emission is highly dependent on
polarization effects such effects should also be taken into account in the
framework of this theory. The basic interaction between a single conduction
band, a single valence band and a single light field polarization are
accurately represented by equations (\ref{eq:bulk}-\ref{eq:endbulk})
and (\ref{eq:QW}-\ref{eq:endQW}).
By adding the contributions of separate transitions, any many band system
may be described based on these equations. In semiconductor quantum wells
the situation is considerably simplified if only the lowest subbands are
considered. Then there are only two completely separate transitions involving
circular light field polarizations coupled to a single one of the two electron and hole
bands. The quantum well structure does not interact with light fields which are
linearly polarized in the direction perpendicular to the plane of the quantum well.
The equations for quantum wells are therefore completed by adding an index of
$+$ or $-$ to each variable.

The situation in the bulk system is much more involved. The transitions
occur between the two fold degenerate spin 1/2 system of the electrons
and the four fold degenerate spin 3/2 system of the holes. All three
polarization directions of the light field are equally possible, connecting
each of the electron bands with three of the four hole bands. However,
since the effective mass of the two heavy hole bands is much
larger than the effective mass of the light holes (e.g.~by a factor of eight in GaAs),
only a small fraction of the holes will be in the light-hole bands (about
6\% in GaAs for equilibrium distributions). Consequently, the carrier
subsystem can again be separated into two pairs of bands. However, the
light field polarization emitted by the electron-heavy hole transitions in
bulk material is circularly polarized with respect to the relative
momentum $2 {\bf k}$ of the electron and the heavy hole. Since usually there
is no strong directional anisotropy in the ${\bf k}$ space distribution of the
carriers, it can be assumed that one third of the ${\bf k}$ space volume
contributes to each polarization direction and the equations may be formulated
accordingly.

In the following, we will assume that the Wigner distributions of the
two pairs of bands considered are approximately equal at all times.
Note that this means that hole burning effects in the spin and polarization
dynamics which may occur in vertical cavity surface emitting lasers
\cite{Mig95}
are ignored. However, such effects have been investigated in several other
studies \cite{Mar95,Hof97,Lem97,Hof98a} and are found to be fairly weak in some
devices \cite{Jan97}.

The complete set of dynamical equations can now be formulated by adding
the carrier dynamics and the linear part of Maxwell's equations to the
light-carrier interaction.


\section{carrier dynamics}
\label{sec:cdyn}
Modeling the carrier dynamics of a semiconductor system can be a formidable
task all by itself.
A number of approximations and models have been developed
to deal with the effects of many-particle interactions and correlations and
with the dissipation caused by the electron-phonon interactions
\cite{Cho94,Kuh92}. In the
following, we choose a simple diffusion model.
Many particle effects such as the band gap renormalization
or the Coulomb enhancement are not mentioned explicitly, but can
be added in a straightforward manner \cite{OH96pqe,Hes96}.

We assume that the electron and hole densities will be kept equal by
the Coulomb interaction, which will induce a current whenever charges are
separated. Therefore, it is possible to define the ambipolar carrier
density $N({\bf r})$ with
\begin{eqnarray}
N({\bf r})   &=& \frac{1}{4\pi^3}\!\int\!d^3{\bf k}f^e({\bf r},{\bf k})
\nonumber \\ &=& \frac{1}{4\pi^3}\!\int\!d^3{\bf k}f^h({\bf r},{\bf k}).
\end{eqnarray}
Note that the two fold degeneracy of the electron and heavy hole bands
has been included by choosing a density of states of $\frac{1}{4\pi^3}$
instead of $\frac{1}{8\pi^3}$. This includes the assumption that the
Wigner distribution does not depend on the spin variable of the electrons
and holes as mentioned above.

The light carrier interaction of this carrier density is
\begin{eqnarray}
\label{eq:trans}
\left.\frac{\partial}{\partial t} N({\bf r}) \right |_{cL} &=&
 i g_0 \frac{\sqrt{\nu_0}}{4\pi^3}
     \int d^3{\bf k}\sum_i \left(C_{ii}({\bf r};{\bf r},{\bf k})
                       -C_{ii}^*({\bf r};{\bf r},{\bf k})\right)
\nonumber \\ &=& -\left.
\frac{\partial}{\partial t} \sum_i I_{ii}({\bf r};{\bf r})\right |_{cL},
\end{eqnarray}
where the index $i$ denotes the component of the light
field or dipole density corresponding to the linear polarization
direction of $i=x,y,z$.
In the case of $C_{ij}({\bf r};{\bf r^\prime},{\bf k})$
the index $i$ refers to the field polarization and the second index $j$
denotes the vector component of the dipole vector.
Equation (\ref{eq:trans}) shows how the field-dipole correlation converts
electron-hole pairs into photons.
The total carrier density dynamics can now be formulated as
\begin{eqnarray}
\frac{\partial}{\partial t} N({\bf r}) &=&
 D_{amb} \Delta N({\bf r}) + j({\bf r}) - \gamma N({\bf r})
\nonumber \\&&+i g_0 \frac{\sqrt{\nu_0}}{4\pi^3}
     \int d^3{\bf k}\sum_i \left(C_{ii}({\bf r};{\bf r},{\bf k})
                       -C_{ii}^*({\bf r};{\bf r},{\bf k})\right),
\end{eqnarray}
where $D_{amb}$ is the ambipolar diffusion constant, $j({\bf r})$ is the
injection current density, and $\gamma$ is the rate of spontaneous
recombinations by non-radiative processes and/or spontaneous emission into
modes not considered in $I_{ij}({\bf r},{\bf r^\prime})$, e.g. if the paraxial
approximation is applied.

The ${\bf k}$ dependence of the distribution functions
$f^e({\bf r},{\bf k})$
and $f^h({\bf r},{\bf k})$ may be approximated by assuming
that the electrons and holes will always be in thermal equilibrium.
The distribution functions are then given by Fermi functions
\begin{equation}
f_{eq}^{e,h}({\bf r},{\bf k})
=
\left( \exp \left[\frac{1}{k_B T} 
                  \left( \frac{\hbar^2{\bf k}^2}
                              {2m_{eff}^{e,h}} -\mu^{e,h}({\bf r}) \right)
            \right ]+1
\right)^{-1},
\end{equation}
where $m_{eff}^{e,h}$ are the effective masses of electrons and heavy holes,
respectively. The chemical potential $\mu^{e,h}({\bf r})$ is a function
of the carrier density $N({\bf r})$. A useful estimate of this relationship
is given by the Pade approximation \cite{Hes96,Ell89}.
Spectral holeburning may be taken into account by introducing a relaxation
time $\tau_r$ and converting the dynamics of the distribution function
due to the light-carrier interaction into a deviation from the equilibrium
distribution by adiabatic elimination of the relaxation dynamics:
\begin{equation}
f^{e,h}({\bf r},{\bf k})=f_{eq}^{e,h}({\bf r},{\bf k})
+i g_0\sqrt{\nu_0}\tau_r
\sum_i \left(C_{ii}({\bf r};{\bf r},\pm{\bf k})
     -C_{ii}^*({\bf r};{\bf r},\pm{\bf k})\right).
\end{equation}

Finally, the carrier dynamics of the dipole $p({\bf r},{\bf k})$ and the
dipole part of the field-dipole correlation $C({\bf r};{\bf r^\prime},
{\bf k})$ needs to be formulated. Since both depend on a correlation of the
electrons with the holes, they will necessarily relax rather quickly
at a rate of $\Gamma({\bf k})$ which should be of the same order of
magnitude as
$1/\tau_r$. Physically, $\Gamma({\bf k})$ may be interpreted as the total
momentum dependent scattering rate in the carrier system.
The remainder of the dynamics can be derived from the single particle
dynamics. This unitary contribution to the evolution of the dipole
may be expressed by a momentum dependent frequency $\Omega({\bf k})$.
For parabolic bands and isotropic effective masses $m_{eff}^{e/h}$,
this frequency is given by
\begin{equation}
\label{eq:bandshape}
\Omega({\bf k})= \left(\frac{\hbar}{2m_{eff}^e}+
                       \frac{\hbar}{2m_{eff}^h}\right){\bf k}^2.
\end{equation}
Many-particle effects due to the Coulomb interaction between the carriers
may be included by introducing a carrier density dependence in
$\Gamma({\bf k},N({\bf r}))$ and $\Omega({\bf k},N({\bf r}))$. Such
renormalization terms representing the mean field effects of the
carrier-carrier interaction have been derived and discussed e.g. in
\cite{Hes96}. In the following, this many particle renormalization
will not be mentioned explicitly, although it can be included in a
straightforward manner.

With the rates $\Gamma({\bf k})$ and $\Omega({\bf k})$
the dipole dynamics reads
\begin{mathletters}
\begin{eqnarray}
\left.\frac{\partial}{\partial t} C_{ij}({\bf r};{\bf r^\prime},{\bf k})
\right |_{c} &=&
          - \left(\Gamma ({\bf k})+i\Omega({\bf k})\right)
                  C_{ij}({\bf r};{\bf r^\prime},{\bf k})
\\
\left.\frac{\partial}{\partial t} p_{i}({\bf r},{\bf k}) \right |_{c} &=&
          - \left(\Gamma ({\bf k})+i\Omega({\bf k})\right)
                  p_{i}({\bf r},{\bf k}).
\end{eqnarray}
\end{mathletters}
Note that the phase dynamics is formulated relative to the band gap frequency
$\omega_0$. The real physical phase oscillations of $p({\bf r},{\bf k})$
would include an additional phase factor of $\exp[-i\omega_0 t]$. However,
the only physical effect of this oscillation is to establish resonance with
the corresponding frequency range in the electromagnetic field, the dynamics
of which we consider next.


\section{Maxwell's equation}
\label{sec:Max}

The Heisenberg equations of motion describing the operator dynamics of the
electromagnetic field operators are identical to the classical Maxwell's
equations. In terms of the electromagnetic field ${\bf E}
({\bf r})$ and the dipole densities ${\bf P}({\bf r})$, the equation
reads
\begin{equation}
\nabla\times(\nabla\times {\bf E}({\bf r}))
+\frac{\epsilon_r}{c^2}\frac{\partial^2}{\partial t^2}
\left({\bf E}({\bf r})
+\frac{1}{\epsilon_r\epsilon_0}{\bf P}({\bf r})\right)=0,
\end{equation}
where $\epsilon_0$ and $c$ are the dielectric constant and the speed of
light in vacuum, respectively, and $\epsilon_r\epsilon_0$ is the
dielectric constant in the background semiconductor medium.

Maxwell's equation describes the light field dynamics for all frequencies.
Since we are only interested in
frequencies near the band gap frequency $\omega_0$, it is useful to
separate the phase factor of $\exp[-i\omega_0 t]$, defining
${\bf E}({\bf r}) = \exp[-i\omega_0 t]{\bf E_0}({\bf r})$. Now
${\bf E}_0({\bf r})$ can be considered to vary slowly in time relative to
$\exp[-i\omega_0 t]$. Therefore, the time derivatives may be approximated
by
\begin{equation}
\exp[i\omega_0 t]
\frac{\partial^2}{\partial t^2}\exp[-i\omega_0 t]{\bf E}_0({\bf r})
\approx
-\omega_0^2 {\bf E}_0({\bf r})
- 2i\omega_0 \frac{\partial}{\partial t} {\bf E}_0({\bf r}).
\end{equation}
Similarly, ${\bf P}_0({\bf r})$ may be defined such that ${\bf P}({\bf r})
=\exp[-i\omega_0 t]{\bf P}_0({\bf r})$. The approximation used here
may even be of zero order, since we are primarily interested in the dynamics
of the electromagnetic field:
\begin{equation}
\exp[i\omega_0 t]
\frac{\partial^2}{\partial t^2}\exp[-i\omega_0 t]{\bf P}_0({\bf r})
\approx
-\omega_0^2 {\bf P}_0({\bf r}).
\end{equation}
The temporal evolution of the electromagnetic field now reads
\begin{equation}
\label{eq:edynparax1}
\frac{\partial}{\partial t}{\bf E}_0({\bf r}) =
-i \frac{\omega_0}{2k_0^2\epsilon_r}
\left(\nabla\times(\nabla\times{\bf  E}_0({\bf r}))
      - \epsilon_r k_0^2 {\bf E}_0({\bf r})\right)
-i \frac{\omega_0}{2\epsilon_r\epsilon_0}{\bf P}_0({\bf r}),
\end{equation}
where $k_0=\omega_0/c$ is the vacuum wavevector length corresponding to
$\omega_0$. In (\ref{eq:edynparax1}), the field dynamics 
is described in terms of electromagnetic
units, that is the fields represent forces acting on charges. To switch
scales to the photon densities represented by ${\bf\cal E} ({\bf r})$,
energy densities have to be considered.
Since the energy of each
photon will be close to the bandgap energy $\hbar\omega_0$, the energy density
of the electromagnetic field is given by
\begin{equation}
\hbar \omega_0 {\bf\cal E}^*({\bf r}){\bf\cal E}({\bf r}) =
\frac{\epsilon_r\epsilon_0}{2}{\bf E}_0^*({\bf r}){\bf E}_0^*({\bf r}).
\end{equation}
Therefore, the field may be expressed as photon density amplitude using
\begin{equation}
{\bf E}_0({\bf r}) =
\sqrt{\frac{2\hbar\omega_0}{\epsilon_r\epsilon_0}}{\bf\cal E}({\bf r}).
\end{equation}
The dipole density ${\bf P}_0({\bf r})$ may be expressed in terms of
$\rho^{dipole}({\bf r},{\bf r})$ and ${\bf p}({\bf r},{\bf k})$ by noting that
the density $\rho^{dipole}({\bf r},{\bf r})$ is the dipole density in units
of one-half the atomic dipole given by the interband dipole matrix
element ${\bf d}_{cv}$ at $k=0$. The factor of one-half is a logical
consequence
of the property that $\langle \hat{c}_{\bf R}\hat{d}_{\bf R^\prime} \rangle
\leq 1/2$. Thus, a fully polarized lattice would have a dipole density
of $\rho^{dipole}({\bf r},{\bf r}=1)/(2\nu_0)$ which must correspond to
${\bf P}({\bf r})={\bf d}_{cv}/\nu_0$.
Note that ${\bf d}_{cv}$
contains an arbitrary phase factor depending only on the definition of
the states used for its determination. For convenience,
we assume a definition of phases such that ${\bf d}_{cv}$ is real.
The dipole density ${\bf P}_0({\bf r})$ may then be written as
\begin{eqnarray}
{\bf P}({\bf r}) &=& 2 {\bf d}_{cv} \rho^{dipole}
\nonumber \\
&=& \frac{2 \mid{\bf d_{cv}}\mid}{8\pi^3}
\int d^3{\bf k}\; {\bf p}({\bf r},{\bf k}).
\end{eqnarray}
Written in terms of ${\bf \cal E}({\bf r})$ and ${\bf p}({\bf r},{\bf k})$,
the complete field dynamics now reads
\begin{eqnarray}
\frac{\partial}{\partial t} {\bf\cal E}({\bf r}) 
&=& -i\frac{\omega_0}{2 k_0^2\epsilon_r}
\left(\nabla\times(\nabla\times{\bf\cal E}({\bf r})) -
\epsilon_r k_0^2 {\bf \cal E}({\bf r})\right)
-i\sqrt{\frac{\omega_0}{2 \hbar\epsilon_r\epsilon_0}}
\frac{\mid {\bf d_{cv}}\mid}{8\pi^3} \int d^3 {\bf k}\;
{\bf p}({\bf r},{\bf k}) \nonumber \\
&=& -i\frac{\omega_0}{2 k_0^2\epsilon_r}
\left(\nabla\times(\nabla\times{\bf\cal E}({\bf r})) -
\epsilon_r k_0^2 {\bf \cal E}({\bf r})\right)
-ig_0\frac{\sqrt{\nu_0}}{8\pi^3}\int d^3{\bf k}\; {\bf p}({\bf r},{\bf k}).
\end{eqnarray}
The coupling frequency $g_0$ introduced in equation(\ref{eq:hamilton})
may be expressed in terms of the dipole matrix element ${\bf d_{cv}}$:
\begin{equation}
\label{eq:gd}
g_0 = \sqrt{\frac{\omega_0}{2\hbar\epsilon_r\epsilon_0\nu_0}} \;
      \mid {\bf d_{cv}} \mid.
\end{equation}
With this equation, the operator dynamics of the light field operator
$\hat{b}_{\bf R}$ corresponds to the field dynamics of
the Maxwell-Bloch equations for classical fields.
By applying the linear propagation dynamics of the field to $I_{ij}({\bf r};
{\bf r^\prime})$ and $C_{ij}({\bf r};{\bf r^\prime},{\bf k})$ as well, it is
now possible to formulate a complete set of quantum Maxwell-Bloch equations.


\section{Quantum Maxwell-Bloch equations}
\label{sec:QMBE}

\subsection{Quantum Maxwell-Bloch equations for a three dimensional gain medium}

On the basis of the discussion in the previous sections, the quantum Maxwell-Bloch
equations for a bulk gain medium in three dimensions  read
\begin{mathletters}
\begin{eqnarray}
\frac{\partial}{\partial t} N({\bf r}) &=& D_{amb} \Delta N({\bf r})
                          + j ({\bf r}) - \gamma N({\bf r})
\nonumber \\ & &
     +i g_0 \frac{\sqrt{\nu_0}}{8\pi^3}
      \int d^3{\bf k} \sum_i \left(C_{ii}({\bf r};{\bf r},{\bf k})
                          - C_{ii}^*({\bf r};{\bf r},{\bf k})\right)
\\
\frac{\partial}{\partial t} C_{ij}({\bf r};{\bf r^\prime},{\bf k}) &=&
     - \left(\Gamma ({\bf k}) + i\Omega({\bf k})\right)
             C_{ij}({\bf r};{\bf r^\prime},{\bf k})
\nonumber \\ & &
     -i \frac{\omega_0}{2k_0^2}
        \left(\sum_k \frac{\partial}{\partial r_k}\epsilon_r^{-1}
                     \frac{\partial}{\partial r_k}
                     C_{ij}({\bf r};{\bf r^\prime},{\bf k})\right.
\nonumber \\ & & \left.
             -\sum_k \frac{\partial}{\partial r_i}\epsilon_r^{-1}
                     \frac{\partial}{\partial r_k}
                     C_{kj}({\bf r};{\bf r^\prime},{\bf k})
             + k_0^2 C_{ij}({\bf r};{\bf r^\prime},{\bf k}) \right)
\nonumber \\ & &
     +i g_0\frac{2\sqrt{\nu_0}}{3}\;
      \left( f^e_{eq}\left(k;N({\bf r^\prime})\right) 
     + f^h_{eq}\left(k;N({\bf r^\prime})\right) -1 \right)
     I_{ij}({\bf r};{\bf r^\prime})
\nonumber \\ & &
     +i g_0\frac{2\sqrt{\nu_0}}{3}\;\delta({\bf r}-{\bf r^\prime})\delta_{ij}
             f^e_{eq}\left(k;N({\bf r})\right) \cdot f^h_{eq}\left(k;N({\bf r})\right)
\\
\frac{\partial}{\partial t}
       I_{ij}({\bf r};{\bf r^\prime}) &=& - i\frac{\omega_0}{2k_0^2}
   \sum_k\left(\frac{\partial}{\partial r_k}\epsilon_r^{-1}
                     \frac{\partial}{\partial r_k}
              -\frac{\partial}{\partial r^\prime_k}\epsilon_r^{-1}
                     \frac{\partial}{\partial r^\prime_k}\right)
 I_{ij}({\bf r};{\bf r^\prime})
\nonumber \\ & &
+ i\frac{\omega_0}{2k_0^2}\sum_k\left(
               \frac{\partial}{\partial r_i}\epsilon_r^{-1}
                     \frac{\partial}{\partial r_k}
                I_{kj}({\bf r};{\bf r^\prime})
             - \frac{\partial}{\partial r^\prime_j}\epsilon_r^{-1}
                     \frac{\partial}{\partial r^\prime_k}
                I_{ik}({\bf r};{\bf r^\prime})\right)
\nonumber \\ & &
     -i g_0\frac{\sqrt{\nu_0}}{8\pi^3}
         \int d^3{\bf k} \left(C_{ij}({\bf r};{\bf r^\prime},{\bf k})
                              -C_{ji}^*({\bf r^\prime};{\bf r},{\bf k})\right)
\\[1cm]
\frac{\partial}{\partial t}
         p_i({\bf r},{\bf k}) &=&
      - \left(\Gamma ({\bf k}) + i\Omega({\bf k})\right)
              p_i({\bf r},{\bf k})
 \nonumber \\       & & +i g_0\frac{2\sqrt{\nu_0}}{3}\;
              \left( f^e_{eq}\left(k;N({\bf r})\right)
                   + f^h_{eq}\left(k;N({\bf r})\right) -1 \right) {\cal E}_i({\bf r})
\\
\frac{\partial}{\partial t} {\cal E}_i({\bf r}) &=&
       i\frac{\omega_0}{2k_0^2}\left(\sum_k
              \frac{\partial}{\partial r_k}\epsilon_r^{-1}
              \frac{\partial}{\partial r_k} {\cal E}_i({\bf r})
             -\frac{\partial}{\partial r_i}\epsilon_r^{-1}
              \frac{\partial}{\partial r_k}{\cal E}_k({\bf r})
                                    + k_0^2 {\cal E}_i({\bf r})\right)
\nonumber \\      & & - i g_0\frac{\sqrt{\nu_0}}{8\pi^3}
                             \int d^3{\bf k} \; p_i({\bf r},{\bf k}).
\end{eqnarray}
\end{mathletters}
In order to describe a realistic diode one needs to describe not only the
volume of the active region but also the propagation of light out of this
region. This may either be achieved by defining realistic boundary conditions
or by setting all material properties equal to zero outside a finite
active volume and calculating the light field propagation into the outside
medium by varying $\epsilon_r$ in space.

\subsection{Three dimensional quantum Maxwell-Bloch equations for quantum wells}

Next, we will formulate the equations for a quantum well structure. For this
case, we will also describe different cavity structures and the appropriate
paraxial approximations possible for the various types of laser devices.
Using the terminology of equations (\ref{eq:QW}-\ref{eq:endQW}), the
quantum Maxwell-Bloch equations for a multi quantum well structure with Q quantum
wells read
\begin{mathletters}
\begin{eqnarray}
\frac{\partial}{\partial t}
          N({\bf r}_\parallel) &=& D_{amb} \Delta N({\bf r}_\parallel)
                      + j ({\bf r}_\parallel) - \gamma N({\bf r}_\parallel)
\nonumber \\ & &
     +i g_0 \frac{\sqrt{\nu_0}}{4\pi^2}
      \int d^2{\bf k}_\parallel \sum_i
\left(C_{ii_\parallel}({\bf r};{\bf r}_\parallel,{\bf k}_\parallel)_{r_z=z_0}
    - C_{ii_\parallel}^*({\bf r};{\bf r}_\parallel,{\bf k}_\parallel)_{r_z=z_0}
\right)
\\
\frac{\partial}{\partial t}
C_{ij_\parallel}({\bf r};{\bf r^\prime}_\parallel,{\bf k}_\parallel) &=&
 - \left(\Gamma ({\bf k}_\parallel) + i \Omega({\bf k}_\parallel)\right)
   C_{ij_\parallel}({\bf r};{\bf r^\prime}_\parallel,{\bf k}_\parallel)
\nonumber \\ & &
     -i \frac{\omega_0}{2k_0^2}
        \left(\sum_k \frac{\partial}{\partial r_k}\epsilon_r^{-1}
                     \frac{\partial}{\partial r_k}
    C_{ij_\parallel}({\bf r};{\bf r^\prime}_\parallel,{\bf k}_\parallel)\right.
\nonumber \\ & & \left.
             -\sum_k \frac{\partial}{\partial r_i}\epsilon_r^{-1}
                     \frac{\partial}{\partial r_k}
    C_{kj_\parallel}({\bf r};{\bf r^\prime}_\parallel,{\bf k}_\parallel)
+ k_0^2
    C_{ij_\parallel}({\bf r};{\bf r^\prime}_\parallel,{\bf k}_\parallel)
\right)
\nonumber \\ & &
     +i g_0 Q \sqrt{\nu_0}\;
\left(  f^e_{eq}\left(k_\parallel;\frac{N({\bf r^\prime}_\parallel)}{Q}\right)
      + f^h_{eq}\left(k_\parallel;\frac{N({\bf r^\prime}_\parallel)}{Q}\right) -1 \right)
 I_{ij}({\bf r};{\bf r^\prime})_{r_z^\prime=z_0}
\nonumber \\[0.3cm] & &
     +i g_0 Q \sqrt{\nu_0}\;\delta({\bf r}_\parallel-{\bf r^\prime}_\parallel)
      \delta(r_z-z_0) \delta_{ij_\parallel}
      f^e_{eq}\left(k_\parallel;\frac{N({\bf r}_\parallel)}{Q}\right) \cdot
      f^h_{eq}\left(k_\parallel;\frac{N({\bf r}_\parallel)}{Q}\right)
\\
\frac{\partial}{\partial t}
        I_{ij}({\bf r};{\bf r^\prime}) &=& - i\frac{\omega_0}{2k_0^2}
   \sum_k\left(\frac{\partial}{\partial r_k}\epsilon_r^{-1}
                     \frac{\partial}{\partial r_k}
              -\frac{\partial}{\partial r^\prime_k}\epsilon_r^{-1}
                     \frac{\partial}{\partial r^\prime_k}\right)
 I_{ij}({\bf r};{\bf r^\prime})
\nonumber \\ & &
+ i\frac{\omega_0}{2k_0^2}\sum_k\left(
               \frac{\partial}{\partial r_i}\epsilon_r^{-1}
                     \frac{\partial}{\partial r_k}
                I_{kj}({\bf r};{\bf r^\prime})
             - \frac{\partial}{\partial r^\prime_j}\epsilon_r^{-1}
                     \frac{\partial}{\partial r^\prime_k}
                I_{ik}({\bf r};{\bf r^\prime})\right)
\nonumber \\ & &
     -i g_0\frac{\sqrt{\nu_0}}{4\pi^2}
         \int d^2{\bf k}_\parallel
\left(
 C_{ij_\parallel}({\bf r};{\bf r^\prime}_\parallel,{\bf k}_\parallel)
\delta(r_z^\prime-z_0) \right. \nonumber \\ && \left. \hspace{3cm}
-C_{ji_\parallel}^*({\bf r^\prime};{\bf r}_\parallel,{\bf k}_\parallel)
\delta(r_z-z_0)
\right)
\\[1cm]
\frac{\partial}{\partial t}
        p_{i_\parallel}({\bf r}_\parallel,{\bf k}_\parallel)
&=& - \left(\Gamma ({\bf k}_\parallel) + i\Omega({\bf k}_\parallel)\right)
         p_{i_\parallel}({\bf r}_\parallel,{\bf k}_\parallel)
 \nonumber \\       & & +i g_0 Q \sqrt{\nu_0}\;
    \left( f^e_{eq}\left(k_\parallel;\frac{N({\bf r}_\parallel)}{Q}\right)
         + f^h_{eq}\left(k_\parallel;\frac{N({\bf r}_\parallel)}{Q}\right) -1 \right)
 {\cal E}_i({\bf r})_{r_z=z_0}
\\
\frac{\partial}{\partial t} {\cal E}_i({\bf r}) &=&
       i\frac{\omega_0}{2k_0^2}\left(\sum_k
              \frac{\partial}{\partial r_k}\epsilon_r^{-1}
              \frac{\partial}{\partial r_k} {\cal E}_i({\bf r})
             -\frac{\partial}{\partial r_i}\epsilon_r^{-1}
              \frac{\partial}{\partial r_k}{\cal E}_k({\bf r})
                                    + k_0^2 {\cal E}_i({\bf r})\right)
\nonumber \\      & & - i g_0\frac{\sqrt{\nu_0}}{4\pi^2}
                             \int d^2{\bf k}_\parallel \;
     p_{i_\parallel}({\bf r}_\parallel,{\bf k}_\parallel)\delta(r_z-z_0).
\end{eqnarray}
\end{mathletters}
Note that $N({\bf r}_\parallel)$ is the total carrier density. Therefore,
the density per quantum well which determines the chemical potential
of the carrier distribution functions is $N({\bf r}_\parallel)/Q$.
Again, the structure of an external cavity may be considered either by
boundary conditions or by spatially varying $\epsilon_r$. In particular
laser diodes may be described by distinguishing between reflective and
non-reflective edges. If the reflective surface is perpendicular
to the plane of the quantum wells, the laser is an edge emitter. If the
reflectivity is very high on the surface planes parallel to the quantum wells,
the laser is a vertical cavity surface emitting laser (VCSEL).

\subsection{One dimensional quantum Maxwell-Bloch equations for edge
emitting lasers}

In an edge emitting laser, the laser light field propagates in the plane of
the quantum well. Since the $z$-direction is already defined as the one
perpendicular to the quantum well we will define the $y$-direction as the
direction along which the laser light propagates.
A schematic representation of this type of laser geometry is shown in
figure \ref{edge}.
It is possible to
drastically reduce the dimensionality of the equation describing the
edge emitting laser geometry by noting that the
light field polarization of the amplified fields will be in the plane of the
quantum well and
by limiting the analysis to a single longitudinal mode. Effectively, this
corresponds to a light field ${\bf \cal E}({\bf r})$ with the following
properties:
\begin{mathletters}
\begin{eqnarray}
{\cal E}_x ({\bf r}) &:=& {\cal E}_0 (r_x) \xi (r_y,r_z) \\
{\cal E}_z ({\bf r}) &:=& 0 \\
\frac{\partial}{\partial r_y}{\cal E}_y ({\bf r}) &:=&
-\frac{\partial}{\partial r_x}{\cal E}_x ({\bf r}).
\end{eqnarray}
The envelope function $\xi (r_y,r_z)$ describes both the propagation
along the $y$-direction and the confinement along the $z$-direction.
It represents an approximate solution of the wave equation
in the $yz$-plane normalized by
\begin{equation}
\int dr_y\;dr_z\; \mid \xi(r_y,r_z)\mid^2 = 1.
\end{equation}
\end{mathletters}
The equations are then limited to light field modes with the two dimensional
envelope $\xi (r_y,r_z)$. Spontaneous emission into other light field modes
must be considered by including the rate of emission in the carrier
recombination rate $\gamma$. Since the length $L$ of the laser in the
$y$-direction is also an important property of the device, it is included
by considering the openness of the optical cavity. With the reflectivities of the
laser mirrors given by $R_1$ and $R_2$, the light field in the cavity is damped
by losses
through the mirrors at a rate of
\begin{equation}
\label{eq:kappa}
\kappa = - \frac{c}{2L\sqrt{\epsilon_r}} \ln [R_1 R_2].
\end{equation}
The new one-dimensional variables are now defined as follows:
\begin{mathletters}
\begin{eqnarray}
N_{1D} (r_x) &=& \int dr_y N({\bf r}_\parallel) \\
C_0 (r_x;r_x^\prime,{\bf k}_\parallel) &=&
\int dr_y\;dr_z\;dr_y^\prime \xi(r_y,r_z) \xi^*(r_y^\prime,r_z^\prime=z_0)
C_{xx} ({\bf r};{\bf r^\prime}_\parallel,{\bf k}_\parallel) \\
I_0 (r_x;r_x^\prime) &=& \int dr_y\;dr_z\;dr_y^\prime\;dr_z^\prime
\xi(r_y,r_z) \xi^*(r_y^\prime,r_z^\prime)
I_{xx} ({\bf r};{\bf r^\prime}) \\
p_0 (r_x,{\bf k}_\parallel) &=& \int dr_y \xi^*(r_y,r_z=z_0)
p_x ({\bf r}_\parallel, {\bf k}_\parallel) \\
{\cal E}_0 (r_x) &=& \int dr_y\;dr_z \xi^*(r_y,r_z) {\cal E}_x ({\bf r}).
\end{eqnarray}
\end{mathletters}
The carrier density is now given in terms of a one dimensional density.
To obtain the two dimensional carrier density per quantum well, this density
is to be divided by $QL$. Note that the intensity is also given in terms of
photons per unit length.
The dynamics of the edge emitter then reads
\begin{mathletters}
\begin{eqnarray}
\label{eq:1Ddynamics_a}
\frac{\partial}{\partial t} N_{1D}(r_x) &=&
        D_{amb} \frac{\partial^2}{\partial r_x^2}N_{1D}(r_x)
     + L j (r_x) - \gamma N_{1D}(r_x)
\nonumber \\ & &
     +i g_0 \frac{\sqrt{\nu_0}}{4\pi^2}
      \int d^2{\bf k}_\parallel
\left(C_0 (r_x;r_x,{\bf k}_\parallel)
    - C_0^*(r_x;r_x,{\bf k}_\parallel)
\right)
\\
\frac{\partial}{\partial t}
C_0(r_x;r_x^\prime,{\bf k}_\parallel) &=&
 - \left(\Gamma ({\bf k}_\parallel) + i\Omega({\bf k}_\parallel)\right)
          C_0(r_x;r_x^\prime,{\bf k}_\parallel)
\nonumber \\ & & - \kappa C_0(r_x;r_x^\prime,{\bf k}_\parallel)
     -i \frac{\omega_0}{2k_0^2\epsilon_r}
   \frac{\partial^2}{\partial r_x^2}C_0(r_x;r_x^\prime,{\bf k}_\parallel)
\nonumber \\ & &
     +i g_0 \sigma \sqrt{\nu_0}\;
\left(  f^e_{eq}\left(k_\parallel;\frac{N_{1D}(r_x^\prime)}{QL}\right)
      + f^h_{eq}\left(k_\parallel;\frac{N_{1D}(r_x^\prime)}{QL}\right) -1 \right)
 I_0(r_x;r_x^\prime)
\nonumber \\ & &
     +i g_0 \sigma \sqrt{\nu_0}\;\delta(r_x-r_x^\prime)
      f^e_{eq}\left(k_\parallel;\frac{N_{1D}(r_x)}{QL}\right) \cdot
      f^h_{eq}\left(k_\parallel;\frac{N_{1D}(r_x)}{QL}\right)
\\
\frac{\partial}{\partial t} I_0(r_x;r_x^\prime) &=& -2\kappa I_0(r_x;r_x^\prime)
- i\frac{\omega_0}{2k_0^2\epsilon_r}
   \left(\frac{\partial^2}{\partial r_x^2}
        -\frac{\partial^2}{\partial r_x^{\prime 2}}\right)
 I_0(r_x;r_x^\prime)
\nonumber \\ & &
     -i g_0\frac{\sqrt{\nu_0}}{4\pi^2}
         \int d^2{\bf k}_\parallel
\left(
 C_0(r_x;r_x^\prime,{\bf k}_\parallel)
-C_0^*(r_x^\prime;r_x,{\bf k}_\parallel)
\right)
\\[1cm]
\frac{\partial}{\partial t} p_0(r_x,{\bf k}_\parallel)
&=& - \left(\Gamma ({\bf k}_\parallel) + i\Omega({\bf k}_\parallel)\right)
             p_0(r_x,{\bf k}_\parallel)
 \nonumber \\       & & +i g_0 \sigma \sqrt{\nu_0}\;
    \left(  f^e_{eq}\left(k_\parallel;\frac{N_{1D}(r_x)}{QL}\right)
          + f^h_{eq}\left(k_\parallel;\frac{N_{1D}(r_x)}{QL}\right) -1 \right)
 {\cal E}_0(r_x)
\\
\label{eq:1Ddynamics_b}
\frac{\partial}{\partial t} {\cal E}_0(r_x) &=& -\kappa {\cal E}_0(r_x)
      +i\frac{\omega_0}{2k_0^2\epsilon_r}
              \frac{\partial^2}{\partial r_x^2}
              {\cal E}_0(r_x)
      -i g_0\frac{\sqrt{\nu_0}}{4\pi^2}
                             \int d^2{\bf k}_\parallel \;
     p_0(r_x,{\bf k}_\parallel),
\end{eqnarray}
\end{mathletters}
where $\sigma$ is the confinement factor which determines the overlap
between the quantum wells and the light field mode,
\begin{equation}
\label{eq:1Dsigma}
\sigma = Q \int dr_y \mid \xi (r_y,r_z=z_0) \mid^2.
\end{equation}

\subsection{Two dimensional quantum Maxwell-Bloch equations for VCSELs}

In a VCSEL the light field is strongly confined by highly reflective mirrors
above and below the quantum wells. The light field propagates perpendicular
to the quantum well structure as shown in figure \ref{VCSEL}.
Therefore both the possible polarization directions and the spatial
dynamics remain two dimensional. Only the $z$-direction may be eliminated
by averaging over a single longitudinal mode. Coupling terms between the
polarization directions should be taken into account, even if they are small.
For VCSELs, the assumptions read
\begin{mathletters}
\begin{eqnarray}
{\bf \cal E} ({\bf r}) &\approx&
\tilde{\bf \cal E} ({\bf r}_\parallel) \xi (r_z)  \\
\frac{\partial}{\partial r_z}{\cal E}_z ({\bf r}) &\approx&
-\frac{\partial}{\partial r_x}
\left({\cal E}_x ({\bf r})
+\frac{2g_0\sqrt{\nu_0}}{4\pi \omega_0}
\int d^2{\bf k}_\parallel \delta(r_z-z_0)
p_x ({\bf r}_\parallel,{\bf k}_\parallel)\right) \nonumber \\ &&
-\frac{\partial}{\partial r_y}\left({\cal E}_y ({\bf r})
+\frac{2g_0\sqrt{\nu_0}}{4\pi \omega_0}
\int d^2{\bf k}_\parallel \delta(r_z-z_0)
p_y ({\bf r}_\parallel,{\bf k}_\parallel)\right).
\end{eqnarray}
The latter condition takes into account the divergence of
${\bf\cal E}({\bf r})$ caused by the polarization ${\bf p}_\parallel
({\bf r}_\parallel,{\bf k}_\parallel)$. This is an important contribution
to the quantum Maxwell-Bloch equations, since it coherently couples orthogonal
polarizations.
The properties of the envelope function $\xi (r_z)$ are defined as
\begin{eqnarray}
\int dr_z\; \mid \xi(r_z)\mid^2 &=& 1 \\
\frac{\partial^2}{\partial r_z^2} \xi (r_z) &\approx& -k_0^2 \xi(r_z).
\end{eqnarray}
\end{mathletters}
The cavity loss rate $\kappa$ is defined as in equation (\ref{eq:kappa}).
However, the experimentally observed polarization stability is taken into
account by using slightly different reflectivities for the x and y
polarizations. The cavity loss rate is therefore given by $\kappa_x$
and $\kappa_y$. Experimental results \cite{Jan97} suggest that
$(\kappa_y-\kappa_x)/(\kappa_x)\approx 10^{-3}-10^{-2}$.
A birefringence of $\delta\omega_{x/y}$ is also included to denote
the difference between the band gap frequency and the longitudinal
frequencies of the confined light field for the two polarization
directions. The birefringence $\delta\omega_x-\delta\omega_y$ is
usually in the GHz range.
Since all coordinates are two dimensional, the index $_\parallel$
which
marked the two dimensional coordinates in the quantum well equations
will be omitted. Instead, the two dimensional variables are marked with
a tilde. All indices and coordinates of such variables are only defined in
two dimensions. The variables are defined as
\begin{mathletters}
\begin{eqnarray}
\tilde{N} ({\bf r}) &=& N({\bf r}_\parallel) \\
\tilde{C}_{ij} ({\bf r};{\bf r}^\prime,{\bf k}) &=&
\int dr_z \xi(r_z) \xi^*(r_z^\prime=z_0)
C_{ij_\parallel} ({\bf r};{\bf r^\prime}_\parallel,{\bf k}_\parallel) \\
\tilde{I}_{ij} ({\bf r};{\bf r}^\prime) &=&
\int dr_z\;dr_z^\prime
\xi(r_z) \xi^*(r_z^\prime)
I_{ij} ({\bf r};{\bf r^\prime}) \\
\tilde{p}_i ({\bf r},{\bf k}) &=& \xi^*(r_z=z_0)
p_{i_\parallel} ({\bf r}_\parallel, {\bf k}_\parallel) \\
\tilde{\cal E}_i ({\bf r}) &=& \int dr_z \xi^*(r_z) {\cal E}_i ({\bf r}).
\end{eqnarray}
\end{mathletters}
Note that the dipole variable
$p_{i_\parallel} ({\bf r}_\parallel, {\bf k}_\parallel)$ is rescaled
by the field density of the envelope function $\xi (r_z)$ at $r_z=z_0$.
The dipole given by $\tilde{p}_i ({\bf r},{\bf k})$ is therefore the
average dipole density within the whole cavity, not just within the
quantum wells.
The quantum Maxwell-Bloch equations for the VCSEL now read
\begin{mathletters}
\label{eq:qmbevcsel}
\begin{eqnarray}
\frac{\partial}{\partial t} \tilde{N}({\bf r}) &=& D_{amb} \Delta \tilde{N}({\bf r})
                    + \tilde{j} ({\bf r}) - \gamma \tilde{N}({\bf r})
\nonumber \\ & &
     +i g_0 \frac{\sqrt{\nu_0}}{4\pi^2}
      \int d^2{\bf k} \left( \sum_i
     (\tilde{C}_{ii}({\bf r};{\bf r},{\bf k})
    - \tilde{C}_{ii}^*({\bf r};{\bf r},{\bf k})
)\right.
\nonumber \\ & & \left.
+ \sum_{ij}\frac{1}{k_0^2\epsilon_r}
  \int d^2{\bf r^\prime}\delta({\bf r}-{\bf r^\prime})
 \left( \frac{\partial^2}{\partial r^\prime_i r^\prime_j}
  \tilde{C}_{ij}({\bf r};{\bf r^\prime},{\bf k})
- \frac{\partial^2}{\partial r_i r_j}
\tilde{C}_{ij}^*({\bf r^\prime};{\bf r},{\bf k})\right)
\right)
\\
\frac{\partial}{\partial t}
\tilde{C}_{ij}({\bf r};{\bf r^\prime},{\bf k}) &=&
 - \left(\Gamma ({\bf k}) + i\Omega({\bf k})\right)
          \tilde{C}_{ij}({\bf r};{\bf r^\prime},{\bf k})
\nonumber \\ & & -(\kappa_i-i\delta\omega_i)
                  \tilde{C}_{ij}({\bf r};{\bf r^\prime},{\bf k})
     -i \frac{\omega_0}{2k_0^2\epsilon_r} \Delta_{\bf r}
        \tilde{C}_{ij}({\bf r};{\bf r^\prime},{\bf k})
\nonumber \\ & &
     +i g_0 \sigma \sqrt{\nu_0}\;
\left(  f^e_{eq}\left(k;\frac{\tilde{N}({\bf r^\prime})}{Q}\right)
      + f^h_{eq}\left(k;\frac{\tilde{N}({\bf r^\prime})}{Q}\right) -1 \right)
 \tilde{I}_{ij}({\bf r};{\bf r^\prime})
\nonumber \\[0.3cm] & &
     +i g_0 \sigma \sqrt{\nu_0}\;\delta({\bf r}-{\bf r^\prime})\delta_{ij}
      f^e_{eq}\left(k;\frac{\tilde{N}({\bf r})}{Q}\right) \cdot
      f^h_{eq}\left(k;\frac{\tilde{N}({\bf r})}{Q}\right)
\\
\frac{\partial}{\partial t} \tilde{I}_{ij}({\bf r};{\bf r^\prime}) &=&
- (\kappa_i + \kappa_j)\tilde{I}_{ij}({\bf r};{\bf r^\prime}
- i\frac{\omega_0}{2k_0^2\epsilon_r}
   \left(\Delta_{\bf r}-\Delta_{\bf r^\prime}\right)
 \tilde{I}_{ij}({\bf r};{\bf r^\prime})
\nonumber \\ & &
     -i g_0\frac{\sqrt{\nu_0}}{4\pi^2}
         \int d^2{\bf k}
\left(
 \tilde{C}_{ij}({\bf r};{\bf r^\prime},{\bf k})
-\tilde{C}_{ji}^*({\bf r^\prime};{\bf r},{\bf k})\right.
\nonumber \\ &&
+ \sum_k \frac{1}{k_0^2\epsilon_r}\left.\left(
  \frac{\partial^2}{\partial r^\prime_j r^\prime_k}
  \tilde{C}_{ik}({\bf r};{\bf r^\prime},{\bf k})
- \frac{\partial^2}{\partial r_i r_k}
 \tilde{C}_{jk}^*({\bf r^\prime};{\bf r},{\bf k})
\right)\right)
\\[1cm]
\frac{\partial}{\partial t} \tilde{p}_{i}({\bf r},{\bf k})
&=& - \left(\Gamma ({\bf k}) + i\Omega({\bf k})\right)
             \tilde{p}_{i}({\bf r},{\bf k})
 \nonumber \\       & & +i g_0 \sigma \sqrt{\nu_0}\;
    \left(  f^e_{eq}\left(k;\frac{\tilde{N}({\bf r})}{Q}\right)
          + f^h_{eq}\left(k;\frac{\tilde{N}({\bf r})}{Q}\right) - 1 \right)
 \tilde{\cal E}_i({\bf r})
\\
\frac{\partial}{\partial t} \tilde{\cal E}_i({\bf r}) &=&
      -(\kappa_i+i\delta\omega_i) \tilde{\cal E}_i({\bf r})
      +i\frac{\omega_0}{2k_0^2\epsilon_r}\Delta \tilde{\cal E}_i({\bf r})
\nonumber \\ &&
- i g_0\frac{\sqrt{\nu_0}}{4\pi^2}
   \int d^2{\bf k}\;\left(\tilde{p}_{i}({\bf r},{\bf k})
       + \sum_j \frac{1}{k_0^2\epsilon_r}\frac{\partial^2}{\partial r_i r_j}
                          \tilde{p}_{j}({\bf r},{\bf k})\right),
\end{eqnarray}
\end{mathletters}
with $\sigma$ being the confinement factor along the $z$-direction,
\begin{equation}
\label{eq:VCSELsigma}
\sigma = Q \mid \xi(r_z=z_0) \mid ^2.
\end{equation}
Equations (\ref{eq:qmbevcsel}) present a starting point for the study of spatial
polarization patterns and fluctuations in VCSELs. For more realistic
models, it may also be desirable to include a spatial dependence
of the birefringence and the dichroism. Also, nonlinear effects may be
introduced, e.g.~by separating the carrier densities for right and
left circular polarization \cite{Mig95}.

\subsection{Statistical interpretation and two time correlations}

Using the equations presented above, it is now possible to calculate the
emergence of a spatially coherent light field in semiconductor
laser diodes both above and below threshold.
Note that the average light field ${\bf E}({\bf r})$ will remain zero at
all times due to the random phases of spontaneous emission processes.
The (average) spatial coherence of the light field, however, does not
vanish and is fully described by the
non-local field-field correlations ${\bf I}({\bf r};{\bf r}^\prime)$
which emerge due to the propagation and/or amplification of the originally
incoherent local spontaneous emissions.
This emergence of coherence as a concequence of incoherent emissions
has been discussed in a temporal context using nonequilibrium Green's
functions in \cite{Hen96}.

In order to understand the physical implications of
the well known absence of an average light field in lasers,
it should be recalled that all the results of
the quantum Maxwell-Bloch equations represent
averages which have to be interpreted in terms of statistical physics.
For example, the field-field correlation
${\bf I}({\bf r};{\bf r^\prime})$ represents a variance of the probability
distribution with respect to the possible spatial electromagnetic field
values. The coherent field observed
in experimental time-resolved measurements will vary randomly
from measurement to measurement according to this probability distribution.
Indeed, the intensity distribution itself will vary depending on the random
phase interference of the eigenmodes given by
${\bf I}({\bf r};{\bf r^\prime})$. The calculated average spatial intensity
distribution ${\bf I}({\bf r};{\bf r})$ only describes the average near field
pattern, which is likely to be close to but not identical with the one
actually observed.
The fluctuations of the actual intensity distribution
around this average, however, are disregarded as a consequence of the
factorization performed in section \ref{sec:lci}.
Moreover, the fluctuations in the carrier density distribution
induced by spatial holeburning associated with these fluctuations of
the intensity distribution have also been disregarded.

While the average spatial coherence of the light field is fully
described by the quantum Maxwell-Bloch equations,
the average expected temporal coherence has not yet been explicitly
considered.
Indeed, it is not necessary to consider temporal coherence in the
closed sets of quantum Maxwell-Bloch equations given above because all the
information required to obtain the correct emission and absorption rates
are incorporated in the field-dipole correlation
${\bf C}({\bf r};{\bf r^\prime},{\bf k})$. As shown in
section \ref{sec:ase}, the spectra of gain and spontaneous emission are
implicitly given by the coherent dipole dynamics which enters into the
temporal evolution of this field-dipole correlation.
If explicit information about the two time correlations is
desired, however, such correlations may be included in the dynamics
by noting that the temporal evolution of the
two time correlations of the field
${\bf I}({\bf r},t;{\bf r^\prime},t^\prime)$ and the two time correlations
of the field-dipole correlation
${\bf C}({\bf r},t;{\bf r^\prime},{\bf k},t^\prime)$ is equivalent to the
dynamics of the field and dipole expectation values\cite{Wal94}.
While the quantum
Maxwell-Bloch equations for the correlations at $t=t^\prime$ remain
unchanged, the evolution of the two time correlations as a function
of $t^\prime>t$ is then given
by an additional pair of equations which depend on the carrier dynamics
given by the solution of the original system of quantum Maxwell-Bloch
equations as presented above. In the case of VCSELs
these additional equations supplementing equations (\ref{eq:qmbevcsel}) read
\begin{mathletters}
\label{eq:lin}
\begin{eqnarray}
\frac{\partial}{\partial t^\prime}
   \tilde{C}_{ij}({\bf r},t;{\bf r^\prime},{\bf k},t^\prime)
&=& - \left(\Gamma ({\bf k}) + i\Omega({\bf k})\right)
             \tilde{C}_{ij}({\bf r},t;{\bf r^\prime},{\bf k},t^\prime)
 \nonumber \\       & & +i g_0 \sigma \sqrt{\nu_0}\;
    \left(  f^e_{eq}\left(k;\frac{\tilde{N}({\bf r},t^\prime)}{Q}\right)
          + f^h_{eq}\left(k;\frac{\tilde{N}({\bf r},t^\prime)}{Q}\right) - 1
 \right)
 \tilde{I}_{ij}({\bf r},t;{\bf r^\prime},t^\prime)
\\
\frac{\partial}{\partial t^\prime}
\tilde{I}_{ij}({\bf r},t;{\bf r^\prime},t^\prime)
  &=&
      -(\kappa_j+i\delta\omega_j)
         \tilde{I}_{ij}({\bf r},t;{\bf r^\prime},t^\prime)
      +i\frac{\omega_0}{2k_0^2\epsilon_r}\Delta_{\bf r^\prime}
        \tilde{I}_{ij}({\bf r},t;{\bf r^\prime},t^\prime)
\nonumber \\ &&
- i g_0\frac{\sqrt{\nu_0}}{4\pi^2}
   \int d^2{\bf k}\;
      \tilde{C}_{ij}({\bf r},t;{\bf r^\prime},{\bf k},t^\prime)
\nonumber \\ &&
- i g_0\frac{\sqrt{\nu_0}}{4\pi^2}
   \int d^2{\bf k}\;
    \sum_k \frac{1}{k_0^2\epsilon_r}
     \frac{\partial^2}{\partial r^\prime_j r^\prime_k}
      \tilde{C}_{ik}({\bf r},t;{\bf r^\prime},{\bf k},t^\prime).
\end{eqnarray}
\end{mathletters}
If the carrier density changes slowly, the equations describe the linear
response of the medium caused by the initial intensity distribution
$\tilde{I}_{ij}({\bf r},t;{\bf r^\prime},t)$ and the initial field-dipole
correlation $\tilde{C}_{ij}({\bf r},t;{\bf r^\prime},{\bf k},t)$ determined
from equations (\ref{eq:qmbevcsel}). In this case it is possible to derive
the spectrum and the gain from the eigenmodes and the associated eigenvalues
of the quasi-stationary linear optical system. Note that the eigenmodes
of the linearized dynamics are not necessarily identical with the
eigenmodes of the
intensity distribution $\tilde{I}_{ij}({\bf r},t;{\bf r^\prime},t)$,
since fast variations in the carrier density distribution may have induced
phase locking between the dynamical eigenmodes.

In general, the frequency spectra of the light field are given by
Fourier transforms of the two-time correlations obtained from
equations (\ref{eq:lin}).
However, as noted above, such spectra do not comprise any effects which
arise from carrier density fluctuations. These effects are known
to be quite significant. Well known examples of carrier fluctuation effects
in the frequency spectrum of semiconductor lasers are the 
linewidth enhancement phenomenologically described by the 
linewidth enhancement factor $\alpha$ and the
relaxation oscillation sidebands observed in stable single mode
operation. Moreover, spatial carrier density fluctuations
may also significantly modify the
multi mode spectra of semiconductor laser devices, as pointed out
recently for the case of semiconductor laser arrays \cite{Hof98b}.

\section{Amplified spontaneous emission properties: analytical
results of the quantum Maxwell-Bloch equations}
\label{sec:ase}

\subsection{Gain and spontaneous emission}

For a given carrier distribution $f^{e/h}(k)$, the dynamics of the optical
field ${\bf \cal E}({\bf r})$ and the dipole density ${\bf p}({\bf r})$
are linear. In this case, it is possible to integrate the equations of motion
to obtain a Green's function for the field dynamics.
The equation for bulk material reads
\begin{mathletters}
\begin{eqnarray}
\left.\frac{\partial}{\partial t}{\cal E} ({\bf r},t)\right |_{cL}
& = & g_0^2\frac{\nu_0}{12\pi^3}\int d^3{\bf k}\;
      \int_0^\infty d\tau\; e^{-(\Gamma({\bf k})+i\Omega({\bf k}))\tau}
\nonumber \\
&& \times
\left( f^e(k)+f^h(k)-1 \right)
{\cal E}({\bf r},t-\tau).
\end{eqnarray}
Correspondingly, the equation for a multi quantum well 
structure of $Q$ quantum wells has the form
\begin{eqnarray}
\left.\frac{\partial}{\partial t}{\cal E}({\bf r},t)\right |_{cL} &=&
g_0^2\frac{Q\nu_0}{4\pi^2}\delta(r_z-z_0)\int d^2{\bf k}_\parallel\;
\int_0^\infty d\tau\;
e^{-(\Gamma({\bf k}_\parallel)+i\Omega({\bf k}_\parallel))\tau}
\nonumber \\ && \times
\left( f^e(k_\parallel)+f^h(k_\parallel)-1 \right)
{\cal E}({\bf r},t-\tau).
\end{eqnarray}
\end{mathletters}
An expression for the rate $G(\omega)$ at which a light 
field mode of frequency $\omega$ is amplified
can be derived by solving the integral over $\tau$ using
$
{\cal E}({\bf r},t-\tau)\approx e^{i\omega\tau}{\cal E}({\bf r},t)
$.
The real part of the result is
the gain spectrum given in terms of amplification per unit time, $G(\omega)$.
For bulk material, this amplification rate is given by
\begin{mathletters}
\begin{equation}
\label{eq:bulkgain}
G_{bulk}(\omega)=
g_0^2\frac{\nu_0}{12\pi^3}\int d^3{\bf k}\;
\frac{\Gamma({\bf k})}
     {\Gamma^2({\bf k})+(\Omega({\bf k})-\omega)^2}
\left( f^e(k)+f^h(k)-1 \right),
\end{equation}
and for quantum wells, the corresponding amplification rate reads
\begin{equation}
\label{eq:qwgain}
G_{QW}(\omega)=
g_0^2\frac{Q\nu_0}{4\pi^2}\delta(r_z-z_0)\int d^2{\bf k}_\parallel\;
\frac{\Gamma({\bf k}_\parallel)}
     {\Gamma^2({\bf k}_\parallel)+(\Omega({\bf k}_\parallel)-\omega)^2}
\left( f^e(k_\parallel)+f^h(k_\parallel)-1 \right).
\end{equation}
\end{mathletters}
The gain per unit length can be obtained by dividing the rate $G(\omega)$
by the speed of light in the semiconductor medium, $c\epsilon_r^{-1/2}$.
However, in order to establish the connection between the gain spectrum
and the spectral density of spontaneous emission, it is more convenient to
use the amplification rate as a starting point.

The quantum Maxwell-Bloch equations for the field-dipole correlation
$C_{ij}({\bf r};{\bf r}^\prime,{\bf k})$ show that the ratio between
the spontaneous contributions and the stimulated contributions is
\begin{equation}
\frac{\left. \frac{\partial}{\partial t}
             C_{ij}({\bf r};{\bf r}^\prime,{\bf k})\right|_{spontaneous}}
     {\left. \frac{\partial}{\partial t}
             C_{ij}({\bf r};{\bf r}^\prime,{\bf k})\right|_{stimulated}}
= \frac{\delta({\bf r}-{\bf r}^\prime)\delta_{ij}f^e(k)\cdot f^h(k)}
       {I_{ij}({\bf r};{\bf r}^\prime)\left(f^e(k)+f^h(k)-1\right)}.
\end{equation}
In this equation, $\delta({\bf r}-{\bf r}^\prime)\delta_{ij}$ corresponds
to a photon density of one photon per mode. The spectral density of
the spontaneous emission may therefore be obtained by replacing
$(f^e(k)+f^h(k)-1)$ with $f^e(k)\cdot f^h(k)$ in equations (\ref{eq:bulkgain})
and (\ref{eq:qwgain}), respectively, and multiplying the resulting rates
with twice the density of light field modes which couple to the medium
(the factor of two being a result of considering intensities instead of fields).
At the band edge frequency $\omega_0$, the density of light field modes
per volume and frequency interval in a continuous medium is
\begin{equation}
\rho_{light}=\frac{\omega_0^2}{\pi^2c^3}\epsilon_r^{3/2}.
\end{equation}
The density of the spontaneous emission rate for bulk material
$S_{bulk}(\omega)$ thus reads
\begin{mathletters}
\begin{equation}
\label{eq:bulkspont}
S_{bulk}(\omega)=
\rho_{light} g_0^2\frac{\nu_0}{6\pi^3}\int d^3{\bf k}\;
\frac{\Gamma({\bf k})}
     {\Gamma^2({\bf k})+(\Omega({\bf k})-\omega)^2}
     f^e(k)\cdot f^h(k).
\end{equation}
For quantum wells, the density of modes is only $2/3$ of $\rho_{light}$,
since the dipole component perpendicular to the quantum well is zero.
Also, the delta function $\delta(r_z-z_0)$ may be omitted to obtain
the emission density per area.
The spontaneous emission density $S_{QW}(\omega)$ is then given by
\begin{equation}
\label{eq:qwspont}
S_{QW}(\omega)=
\frac{2}{3}\rho_{light} g_0^2\frac{Q\nu_0}{2\pi^2}\delta(r_z-z_0)
\int d^2{\bf k}_\parallel\;
\frac{\Gamma({\bf k}_\parallel)}
     {\Gamma^2({\bf k}_\parallel)+(\Omega({\bf k}_\parallel)-\omega)^2}
     f^e(k_\parallel)\cdot f^h(k_\parallel).
\end{equation}
\end{mathletters}
The total rate of spontaneous emission per unit volume or area may be
obtained by integrating over all frequencies. This integral removes
the dependence on $\Gamma({\bf k})$. For bulk material,
\begin{mathletters}
\begin{equation}
\int d\omega\;  S_{bulk}(\omega)=
\rho_{light}\; g_0^2\frac{\nu_0}{6\pi^2}\int d^3{\bf k}\;
     f^e(k)\cdot f^h(k)
\end{equation}
and for quantum wells,
\begin{equation}
\int d\omega\; S_{QW}(\omega)=
\rho_{light}\; g_0^2\frac{Q\nu_0}{3\pi}\delta(r_z-z_0)
\int d^2{\bf k}_\parallel\;
     f^e(k_\parallel)\cdot f^h(k_\parallel).
\end{equation}
\end{mathletters}
For zero temperature, the spontaneous emission rate may be derived
by noting that $f^e(k)\cdot f^h(k) = f^e(k) = f^h(k)$. The integral
over ${\bf k}$ may thus be solved, resulting in
\begin{mathletters}
\begin{equation}
\int d^3{\bf k}\; f^e(k)\cdot f^h(k) = 4\pi^3 N
\end{equation}
for bulk material and
\begin{equation}
\int d^2{\bf k}_\parallel\; f^e(k_\parallel)\cdot f^h(k_\parallel) =
\frac{2\pi^2}{Q} N
\end{equation}
\end{mathletters}
for quantum wells. For both bulk and quantum wells the density of
spontaneous emission now may be expressed as
\begin{equation}
\label{eq:stot}
S_{total}=\frac{N}{\tau_s},
\end{equation}
where $1/\tau_s$ is the rate of spontaneous emission given by
\begin{equation}
\label{eq:tauspont}
\frac{1}{\tau_s}= \frac{2\pi}{3} \rho_{light}g_0^2\nu_0.
\end{equation}
Using equation (\ref{eq:gd}), the rate of spontaneous emission may
also be expressed in terms of the dipole matrix element $d_{cv}$,
\begin{eqnarray}
\frac{1}{\tau_s} &=& \frac{4}{\hbar\omega_0}
                     \epsilon_r^{1/2} \frac{1}{4\pi\epsilon_0}
                     \frac{\omega_0^4}{3c^3}\mid d_{cv}\mid^2
\nonumber \\     &=& \frac{4}{\hbar\omega_0} P_{rad},
\end{eqnarray}
where $P_{rad}$ is the classical power radiated by an oscillating dipole
of the amplitude $d_{cv}$. Note that the factor of $4/\hbar\omega_0$ is
consistent with
a quantum noise interpretation of spontaneous emission such as the one
represented by the semiclassical Langevin equations. According to this
interpretation, the fluctuations of each dipole are given by
$2 \mid d_{cv}\mid^2$ because both the real and the imaginary part of
the dipole contribute. The spontaneous emission of an excited atom is
then composed of one half amplified field noise and one half dipole
fluctuations. Therefore, each one of the two dipole components contributes
one quarter of the total spontaneous emission of an excited state.
Numerical values of $\mid {\bf d}_{cv} \mid$ and $g_0$ for GaAs
may be determined by assuming a spontaneous lifetime of $\tau_s=3$~ns,
a band gap of $\hbar\omega_0 = 1.5$ eV and $\epsilon_r=12$. The dipole matrix
element is then $\mid {\bf d}_{cv} \mid = 4.3\times 10^{-29} Cm$
which corresponds to a distance of $2.7\times 10^{-10}$m times the electron
charge and the coupling frequency is $g_0 = 2.1\times 10^{15}\mbox{s}^{-1}$.

For quantum wells at zero temperature, the integrals over ${\bf k}_\parallel$
can be solved analytically using equation (\ref{eq:bandshape}) and assuming
that $\Gamma$ is independent of ${\bf k}_\parallel$.
The resulting gain spectrum is then given by
\begin{equation}
\label{eq:wqw}
\frac{\epsilon_r^{1/2}}{c}G_{QW}(\omega)=\frac{\epsilon_r^{1/2}}{c}
g_0^2\frac{Q\nu_0}{2\pi\hbar}\frac{m_{eff}^e m_{eff}^h}{m_{eff}^e+m_{eff}^h}
\delta(r_z-z_0)\left( 2 \arctan \left(\frac{\Omega_f-\omega}{\Gamma}\right)
+  \arctan \left(\frac{\omega}{\Gamma}\right) - \frac{\pi}{2}\right),
\end{equation}
where $\Omega_f$ is the transition frequency at the Fermi surface of the
electrons and holes. It is related to the carrier density $N$ by
\begin{equation}
\label{eq:fermif}
\Omega_f = \frac{\pi\hbar}{Q}
           \frac{m_{eff}^e+m_{eff}^h}{m_{eff}^e m_{eff}^h} N.
\end{equation}
The spectral density of spontaneous emission is thus given by
\begin{equation}
\label{eq:sqw}
S_{QW}(\omega)=
\rho_{light} g_0^2\frac{2 Q\nu_0}{3\pi\hbar}
\frac{m_{eff}^e m_{eff}^h}{m_{eff}^e+m_{eff}^h}
\left(\arctan \left(\frac{\Omega_f-\omega}{\Gamma}\right)
+  \arctan \left(\frac{\omega}{\Gamma}\right)\right).
\end{equation}

Typical spectra of gain and spontaneous emission of an active layer
containing $Q=5$ quantum wells obtained from the analytic approximations
(\ref{eq:wqw}) and (\ref{eq:sqw}), respectively, are presented in Fig.~\ref{spectra}
for characteristic values of the carrier density.
For the spectra in Fig.~\ref{spectra} we have assumed a
total spontaneous emission lifetime of $\tau_s=3$ ns and a band gap of
$\hbar\omega_0=1.5$ eV. Other parameters are the
effective mass of electrons ($m_{eff}^e = 0.067 m_0$) and holes ($m_{eff}^h= 0.053 m_0$),
given in units of the electron mass $m_0$ as well as the dipole damping rate
$\hbar\Gamma = 8$ meV, and dielectric constant $\epsilon_r=12$. Gain and spontaneous
emission are both displayed relative to the peak values
$S_{max}=2.7\times 10^7 \mbox{cm}^{-2}$ and
$G_{max}= 2.5\times 10^{7} \mbox{cm}\mbox{s}^{-1} \delta(r_z-z_0)$.
The gain value may be interpreted by calculating the gain of a light beam
of width $\sigma^{-1}=10^{-5}$~cm with an incidence perpendicular with respect
to the quantum well and traveling at a speed of $10^{10} \mbox{cm}\mbox{s}^{-1}$ in the plane
of the quantum well. The maximal gain is then given by $250 \mbox{cm}^{-1}$.
The five spectra displayed in Fig.~\ref{spectra} are the gain at $N=0$ ($g_0$), the gain at
$N=10^{12}\mbox{cm}^{-2}$ ($g_1$), the spontaneous emission at
$N=10^{12}\mbox{cm}^{-2}$ ($s_1$), the gain at
$N=5 \times10^{12}\mbox{cm}^{-2}$    ($g_5$)
and the spontaneous emission at
$N=5\times 10^{12}\mbox{cm}^{-2}$    ($s_5$).
Fig.~\ref{spectra} clearly shows the influence of the carrier density $N$.
In the absence of charge carriers ($g_0$), the laser is purely absorptive.
With increasing carrier density $N$, transparency is reached at $N=10^{12}\mbox{cm}^{-2}$ ($g_1$).
This density is characterized by vanishing gain (i.e.~there is neither gain nor absorption)
at a frequency of $\hbar\omega = 0$. At the same time, however, there is a significant
contribution of spontaneous emission  ($s_1$) with a maximum at a frequency of $\hbar\omega \approx 4$~meV.
Finally, at high values of the carrier density ($N=5\times 10^{12}\mbox{cm}^{-2}$),
both gain ($g_5$) and spontaneous emission ($s_5$) have a maximum above the band-gap frequency.

\subsection{Spontaneous emission factor and far field pattern
of an edge emitting laser}

In the optical cavity of a laser the equations for gain and spontaneous
emission are modified by the mode structure. In particular, the total
linear response of an electromagnetic field mode inside the cavity
includes the cavity loss rate $\kappa$.
For the edge emitting semiconductor laser, the gain function of the cavity modes is given by
\begin{equation}
\label{eq:1Dgain}
G_{1D}(\omega)=
g_0^2\frac{\nu_0}{4\pi^2}\sigma\int d^2{\bf k}_\parallel\;
\frac{\Gamma({\bf k}_\parallel)+\kappa}
   {(\Gamma({\bf k}_\parallel)+\kappa)^2+(\Omega({\bf k}_\parallel)-\omega)^2}
\left( f^e(k_\parallel)+f^h(k_\parallel)-1 \right),
\end{equation}
where the confinement factor $\sigma$ is defined according to equation
(\ref{eq:1Dsigma}).
Spontaneous emission into the cavity modes passes
through the gain medium and is thereby absorbed or amplified accordingly.
Thus for an edge emitting laser, the rate of spontaneous emission into a cavity mode of frequency $\omega$
is given by
\begin{equation}
\label{eq:1Dspont}
S_{1D}(\omega)=
2 g_0^2\frac{\nu_0}{4\pi^2}\sigma\int d^2{\bf k}_\parallel\;
\frac{\Gamma({\bf k}_\parallel)+\kappa}
   {(\Gamma({\bf k}_\parallel)+\kappa)^2+(\Omega({\bf k}_\parallel)-\omega)^2}
f^e(k_\parallel) \cdot f^h(k_\parallel).
\end{equation}
This rate represents the total rate of spontaneous emission events per mode,
regardless of the actual width of the laser. On the other hand, equation
(\ref{eq:stot}) gives the total rate of spontaneous emission per quantum
well area.

\subsubsection{Spontaneous emission factor}

In a laser of length $L$ and total width $W$ (c.f.~Fig.~\ref{edge}), the spontaneous
emission rate into free space is $L W\: S_{total}$.
The spontaneous emission factor $\beta$, which is generally
defined as the fraction of spontaneous emission being emitted into the cavity mode
\cite{Ebe93}, is on the basis of our theory given by the expression
\begin{equation}
\beta (\omega,N_{1D})=\frac{\tau_s S_{1D}}{W\; N_{1D}}.
\end{equation}
Note that the two dimensional carrier density $N$ in the quantum well
is related to the one dimensional carrier density $N_{1D}$ by
$N_{1D}=L\; N$.
The spontaneous emission factor $\beta$ is a function of
both frequency and carrier density. Consequently, the common
assumption of the spontaneous emission factor $\beta$ being independent of
the carrier density \cite{Ebe93,Gun94} may be regarded as an approximation
similar to the assumption of linear gain.

An analytical expression for the spontaneous emission factor may be obtained for
zero temperature. The k-space integrals may then be solved analytically
using equation (\ref{eq:bandshape}) and assuming that $\Gamma$
is independent of ${\bf k}_\parallel$.
For the spontaneous emission factor $\beta$, the analytical result reads
\begin{equation}
\label{eq:beta}
\beta(\omega,\Omega_f) =
\frac{3\sigma}{2\pi\rho_{light}Q\; W\; L\; \Omega_f}
\left(\arctan \left(\frac{\Omega_f-\omega}{\Gamma+\kappa}\right)
+  \arctan \left(\frac{\omega}{\Gamma+\kappa}\right)\right).
\end{equation}
The Fermi frequency $\Omega_f$ is defined by equation(\ref{eq:fermif}) and
expresses, in particular, the carrier density dependence of $\beta$.
For $\Omega_f,\omega\ll \Gamma+\kappa$ we recover the result typically given in the
literature (e.g. \cite{Ebe93}) which does not depend on $N$.
Fig.~\ref{beta} shows the deviation of the spontaneous emission factor from this value
as the Fermi frequency $\Omega_f$ passes the point of resonance with the cavity mode.
Fig.~\ref{beta} illustrates, in particular, the carrier density dependence of
$\beta$ for three modes with frequencies above the band gap frequency given by
(a)~$\omega= 0$,
(b)~$\omega= 0.5 (\Gamma+\kappa)$, and
(c) $\omega= \Gamma+\kappa$.
Most notably, $\beta$ is always smaller than the usual
estimate given by $\beta(\omega=\Omega_f=0)$, which
is based on the assumption of ideal resonance between
the transition frequency and the cavity mode.

\subsubsection{Far-field pattern of a broad area laser}

With equations (\ref{eq:1Dgain}) and (\ref{eq:1Dspont}), it is possible to
find the steady state intensity $I_s$ of a mode with frequency $\omega$,
\begin{equation}
I_s(\omega) = \frac{S_{1D}(\omega)}
                   {2\kappa-2G_{1D}(\omega)}.
\end{equation}
Note that this result may also be obtained directly from equations
(\ref{eq:1Ddynamics_a}-\ref{eq:1Ddynamics_b}).
In a wide cavity, the cavity modes are approximately plane wave modes and the relation
between $I_0(\omega)$ and $I_0(r_x;r_x^\prime)$ in that case reads
\begin{equation}
\label{eq:I0}
I_0(r_x,r_x^\prime)=\int dq \; e^{iq(r_x-r_x^\prime)} \;
I \left(\omega=\frac{\omega_0q^2}{2k_0^2\epsilon_r}\right).
\end{equation}
The steady state intensity distribution is characterized by the
spatial coherence derived from the intensity distribution
of the plane wave modes of the cavity. Generally, the intensity
distribution of plane waves corresponds in the far field to an
optical field at angles relative
to the axis of emission in the plane of the quantum well.
The angular distribution of intensity and coherence in the far field is thus given
by
\begin{equation}
I_f(\Theta,\Theta^\prime) = \frac{k_0}{2\pi} \int dr_x dr_x^\prime
\; e^{ik_0\Theta r_x} I_0(r_x,r_x^\prime)e^{-ik_0\Theta^\prime r_x^\prime}.
\end{equation}
Therefore, the far field intensity distribution may be determined
directly from the frequency dependence of the intensities by
\begin{equation}
I_f(\Theta,\Theta) = W k_0 \;
            I_s\left(\omega=\frac{\omega_0\Theta^2}{2\epsilon_r}\right),
\end{equation}
the intensity is given in units of $2\kappa\hbar\omega_0$ per unit angle.

For $T=0$ we may in analogy to (\ref{eq:beta}) solve the integral in (\ref{eq:1Dspont})
by assuming $\Gamma$ to be independent of ${\bf k}$.
As an analytical expression we then obtain for the far-field intensity distribution
of a broad area semiconductor laser
\begin{eqnarray}
I_f(\Theta,\Theta) &=& W k_0 \frac{
   \arctan \left(\frac{\Omega_f-\omega(\Theta)}{\Gamma+\kappa}\right)
+  \arctan \left(\frac{\omega(\Theta)}{\Gamma+\kappa}\right)}
{\pi (R+\frac{1}{2})
-2 \arctan \left(\frac{\Omega_f-\omega(\Theta)}{\Gamma+\kappa}\right)
-  \arctan \left(\frac{\omega(\Theta)}{\Gamma+\kappa}\right)},
\nonumber
\\ \mbox{with} &&
R = \frac{2 \hbar\kappa}{g_0^2\nu_0\sigma}
    \frac{m^e_{eff}+m^h_{eff}}{m^e_{eff} m^h_{eff}}
\nonumber \\ \mbox{and} &&
\omega(\Theta) =\frac{\omega_0\Theta^2}{2\epsilon_r}.
\end{eqnarray}
The parameter $R$ represents the ratio between the cavity loss rate $\kappa$
and the maximum amplification rate of the gain medium. Laser
activity is only possible if $R<1$. The classical laser threshold
is defined by the carrier density for which the denominator of
$I_f(\Theta,\Theta)$ is zero for a single specific frequency
$\omega(\Theta)$. Consequently, the carrier density at which this occurs is
pinned. Figure \ref{farfield} shows the far field intensity distribution
for different carrier densities below this pinning density.
In Fig.~\ref{farfield}~(a), the wide intensity distribution
of amplified spontaneous emission for carrier densities is much lower than
the pinning density. The intensity maximum is clearly located at $\Theta=0$.
Figure \ref{farfield}(a) shows the intensity distribution for carrier
densities halfway towards threshold. Already, the intensity maxima move
to angles of $\pm 15^\circ$, corresponding to the  frequency at which
the gain spectrum has its maximum. In the case of Fig.~\ref{farfield}(c), the
threshold region is very close to the pinning density. The peaks in the far
field pattern narrow as the laser intensity is increased.
Consequently the far field pattern indeed is a measure of the spatial coherence --
similar as the linewidth of the laser spectrum is a measure of temporal coherence.
It is therefore desirable to consider quantum noise effects in
the spatial patterns of optical systems. In the context of squeezing,
such patterns have been investigated by Gatti and coworkers \cite{Gat97}
based on the general formulation of Lugiato and Castelli \cite{Lug92}.
The laser patterns presented here are based on the same principles.
Usually, however, the strong dissipation prevents squeezing in laser systems
unless the pump-noise fluctuations are suppressed \cite{Yam86}.


\section{Conclusions}
\label{sec:concl}

The quantum Maxwell-Bloch equations (QMBE) for spatially inhomogeneous
semiconductor lasers derived in this paper take into account the
quantum mechanical nature of the light field as well as that of the
carrier system. The only approximation used in the derivation of the
intensity and correlation dynamics is that of statistical independence
between the two carrier systems and the light field.
In the QMBE presented here, the effects of coherent spatiotemporal quantum fluctuations
 which are generally not
considered  in the semiclassical Maxwell-Bloch equations for semiconductor laser devices
have thus been taken into account.

The spontaneous emission term appears side by side with the gain and
absorption term in the dynamics of the field-dipole correlation.
In this way the spatial coherence of spontaneous emission and amplified
spontaneous emission is consistently described by the quantum Maxwell-Bloch
equations.
Typical features of the model have been illustrated by the spectra of gain
and spontaneous emission. An example of the spatial coherence characteristics
described by the quantum Maxwell-Bloch equations has been presented by
analytically obtaining the spontaneous emission factor $\beta$
and the far field distribution for the example of a broad area edge
emitting laser.
In general the quantum Maxwell-Bloch equations derived for edge emitting
and vertical cavity surface emitting lasers provide a starting point
for a detailed
analysis of spatial coherence patterns in diverse semiconductor laser
geometries such as broad area or ultra-low threshold lasers.


\begin{appendix}
\section{Stochastic simulation of measurements}
As explained at the end of section \ref{sec:QMBE}, the field average
of zero is an expression of our lack of knowledge about the actual 
physical field present in the laser device. This lack of knowledge
could be removed by performing a measurement on the light field. 
For example, the phase information could be obtained by measuring 
the interference of the light field 
from the diode with light from a separate laser. A simple theoretical
simulation of such a measurement is obtained by assuming that 
${\bf I}({\bf r};{\bf r^\prime})$ defines the variance of a Gaussian 
distribution of coherent states (Gaussian P representation). The measurement
of the actual field ${\cal \bf E}({\bf r})$ may then be simulated by
randomly selecting a coherent field from this Gaussian distribution.
Since the field and the dipole density are correlated, the statistical
selection of ${\cal \bf E}({\bf r})$ shifts the dipole density average
${\bf p}({\bf r},{\bf k})$ from zero to
\begin{equation}
{\bf p}({\bf r},{\bf k}) = 
\frac{\int d{\bf r}^\prime {\cal \bf E}({\bf r})
               {\bf C}({\bf r^\prime};{\bf r},{\bf k})}
     {\sum_{ij}\int d{\bf r}^\prime I_{ij}({\bf r}^\prime;{\bf r}^\prime)}.
\end{equation}
Because the field has been determined with a precision equal to the quantum 
limit, the correlations factorize and the new values of 
${\bf I}({\bf r};{\bf r^\prime})$ and ${\bf C}({\bf r};{\bf r^\prime},
{\bf k})$ are given by the respective products of ${\cal \bf E}({\bf r})$
and ${\bf p}({\bf r},{\bf k})$. 

The drastic changes in the intensity distribution and in the dipole dynamics
which may result from such a simulated measurement cause spatial holeburning
and give rise to relaxation oscillations. By performing several stochastical
simulations starting from the same initial probability distribution, the
statistics of the carrier density fluctuations may be obtained.
It is then possible to derive the correct linewidth enhancement factor
$\alpha$ as well as the relaxation oscillation sidebands directly 
from the dynamics of the quantum Maxwell-Bloch equations.

\section{Langevin equations}
An alternative approach to the problem of quantum noise is given by 
the Langevin equations \cite{Wal94,Gar91}. Langevin equations simulate
quantum noise by adding classical noise sources which reproduce the 
statistical properties given by the uncertainty relations. In
particular, the exponential damping terms given by the cavity loss 
rate $\kappa$ and the dipole relaxation rate $\Gamma$ must be compensated
by a noise input maintaining the quantum fluctuations in the field and dipole
densities. For quantum well lasers, these fluctuations are given by 
\begin{mathletters}
\begin{eqnarray}
\langle {\cal E}^*_i({\bf r}){\cal E}_j({\bf r}) \rangle &=& 
\frac{1}{2} \delta_{ij}\delta({\bf r}-{\bf r^\prime})
\\
\langle p_{i_\parallel}^*({\bf r}_\parallel,{\bf k}_\parallel)
        p_{j_\parallel}({\bf r^\prime}_\parallel,{\bf k^\prime}_\parallel) 
\rangle 
 &=&  \frac{1}{2}\delta_{i_\parallel j_\parallel}
               \delta({\bf r}_\parallel-{\bf r^\prime}_\parallel)
  4\pi^2 \left( f^e_{eq} \left(k_\parallel,N({\bf r_\parallel})\right)\cdot 
                f^h_{eq} \left(k_\parallel,N({\bf r}_\parallel)\right) \right.
\nonumber \\ &&
\qquad \left. {} + \left(1-f^e_{eq}\left(k_\parallel,N({\bf r}_\parallel)\right)\right)\cdot 
                   \left(1-f^h_{eq}\left(k_\parallel,N({\bf r}_\parallel)\right)\right)     \right).
\end{eqnarray}
\end{mathletters}
Note that the variables ${\cal E}_i({\bf r})$ and ${\bf p}({\bf r}_\parallel
,{\bf k}_\parallel)$ 
are classical quantities in this context. The 
Langevin equations presented in the following do not describe quantum 
mechanical coherent states. Instead, semiclassical light fields and dipole
densities are calculated with a precision violating the uncertainty relations.
While the statistics thus derived correspond to the quantum statistics,
the theory applied is fully classical. 

For a VCSEL, the semiclassical Langevin equations corresponding to the 
quantum Maxwell-Bloch equations (\ref{eq:qmbevcsel} read
\begin{mathletters}
\begin{eqnarray}
\label{eq:vcsellangevin}
\frac{\partial}{\partial t} \tilde{N}({\bf r}) &=& D_{amb} \Delta \tilde{N}({\bf r})
                    + \tilde{j} ({\bf r}) - \gamma \tilde{N}({\bf r})
\nonumber \\ & &
     +i g_0 \frac{\sqrt{\nu_0}}{4\pi^2}
      \int d^2{\bf k} \left( \sum_i 
     (\tilde{\cal E}^*_{i}({\bf r})\tilde{p}_i({\bf r},{\bf k})
    - \tilde{\cal E}_{i}({\bf r})\tilde{p}^*_i({\bf r},{\bf k})
)\right.
\nonumber \\ & & \left.
+ \sum_{ij}\frac{1}{k_0^2\epsilon_r}
 \left( \tilde{\cal E}^*_i({\bf r})\frac{\partial^2}{\partial r_i r_j}
  \tilde{p}_{j}({\bf r},{\bf k})
- \tilde{\cal E}_i({\bf r})\frac{\partial^2}{\partial r_i r_j}
\tilde{p}_{j}^*({\bf r},{\bf k})\right)
\right)
\\[0.5cm]
\frac{\partial}{\partial t} \tilde{p}_{i}({\bf r},{\bf k}) 
&=& - \left(\Gamma ({\bf k}) + i\Omega({\bf k})\right) 
             \tilde{p}_{i}({\bf r},{\bf k})
 \nonumber \\       & & +i g_0 \sigma \sqrt{\nu_0}\;
    \left(  f^e_{eq}\left(k;\frac{\tilde{N}({\bf r})}{Q}\right)
          + f^h_{eq}\left(k;\frac{\tilde{N}({\bf r})}{Q}\right) - 1  \right) 
 \tilde{\cal E}_i({\bf r})  + q^p_i ({\bf r},{\bf k},t)
\\
\frac{\partial}{\partial t} \tilde{\cal E}_i({\bf r}) &=& 
      -(\kappa_i+i\delta\omega_i) \tilde{\cal E}_i({\bf r}) 
      +i\frac{\omega_0}{2k_0^2\epsilon_r}\Delta \tilde{\cal E}_i({\bf r})
\nonumber \\ && 
- i g_0\frac{\sqrt{\nu_0}}{4\pi^2}  
   \int d^2{\bf k}\;\left(\tilde{p}_{i}({\bf r},{\bf k})
       + \sum_j \frac{1}{k_0^2\epsilon_r}\frac{\partial^2}{\partial r_i r_j}
                  \tilde{p}_{j}({\bf r},{\bf k})\right) 
      + q^{\cal E}_i({\bf r},t)
\end{eqnarray}
\end{mathletters}
with the noise input terms $q^p_i({\bf r},{\bf k},t)$ and 
$q^{\cal E}_i({\bf r},t)$ given by
\begin{mathletters}
\begin{eqnarray}
\langle {q^p_i}^* ({\bf r},{\bf k},t)
q^p_j ({\bf r^\prime},{\bf k},t+\tau) \rangle &=&
\delta(\tau) \Gamma ({\bf k}) \delta_{ij}
             \delta({\bf r}-{\bf r^\prime})
      4\pi^2 \sigma \left(  f^e_{eq}\left(k,\tilde{N}({\bf r})\right)\cdot 
                            f^h_{eq}\left(k,\tilde{N}({\bf r})\right) \right.
\nonumber \\ &&
\qquad \left. {}  + \left(1-f^e_{eq}\left(k,\tilde{N}({\bf r})\right)\right)\cdot 
                    \left(1-f^h_{eq}\left(k,\tilde{N}({\bf r})\right)\right)      \right)
\\
\langle {q^{\cal E}_i}^*({\bf r},t)q^{\cal E}_i({\bf r}^\prime,t+\tau)\rangle
&=& \delta(\tau)\kappa\delta({\bf r}-{\bf r}^\prime).
\end{eqnarray}
\end{mathletters}
These Langevin equations allow a stochastical simulation of quantum noise
in a semiclassical context. However, it is necessary to perform a large
number of runs before statistical results can be obtained. In particular,
the unrealistic intensity fluctuations of the light field vacuum need to
average out before the results are consistent with quantum theory.
\end{appendix}

%
%
%

%
\begin{figure}
\caption{Schematic representation of the edge emitter geometry.
The laser field is mostly confined to the plane of the quantum well
and propagates along the $y$-axis.}
\label{edge}
\end{figure}
%
%
\begin{figure}
\caption{Schematic representation of a typical VCSEL geometry.
The laser filed propagates perpendicularly to the quantum well along the $z$-axis.
The length of the optical resonator approximately corresponds
to the wavelenth.}
\label{VCSEL}
\end{figure}
%
%
\begin{figure}
\caption{Spectra of gain and spontaneous emission for a quantum well
structure of $Q=5$ quantum wells given relative to
$G_{max}= 2.5\times 10^{7} \mbox{cm}\mbox{s}^{-1} \delta(r_z-z_0)$ and
$S_{max}=2.7\times 10^7 \mbox{cm}^{-2}$.
The five spectra shown are the gain at $N=0$ ($g_0$),
the gain                 at $N=10^{12}\mbox{cm}^{-2}$ ($g_1$),
the spontaneous emission at
$N=10^{12}\mbox{cm}^{-2}$ (s1), the gain at $N=5\times
10^{12}\mbox{cm}^{-2}$ (g5) and the spontaneous emission at $N=5\times
10^{12}\mbox{cm}^{-2}$ (s5).}
\label{spectra}
\end{figure}
%
%
\begin{figure}
\caption{Carrier density dependence of the spontaneous emission
factor $\beta$ for three modes with frequencies above the band gap
frequency given by
(a)~$\omega= 0$,
(b)~$\omega= 0.5 (\Gamma+\kappa)$, and
(c) $\omega= \Gamma+\kappa$.
$\beta_0=\beta(\omega=N=0)$ is determined by the geometry of the laser.
The carrier density is given in terms of the transition frequency at the
Fermi edge $\Omega_f$.}
\label{beta}
\end{figure}
%
%
\begin{figure}
\caption{Far field intensity distributions for $R=0.5$, $\hbar\omega_0=1.5$ eV,
$\hbar(\Gamma+\kappa)=8$ meV and $\epsilon_r=12$. The density is pinned at
$\Omega_f = 1.8805 (\Gamma+\kappa)$. Figure (a) shows the far field
pattern for carrier densities of 0.05, 0.10 and 0.15 times pinning density,
figure (b) shows the distribution for 0.25, 0.5. and 0.75 times pinning
density and figure (c) shows the distribution for 0.90, 0.95 and 0.99 times
pinning density. The peaks appear at emission angles of $\pm 15^\circ$.
}
\label{farfield}
\end{figure}

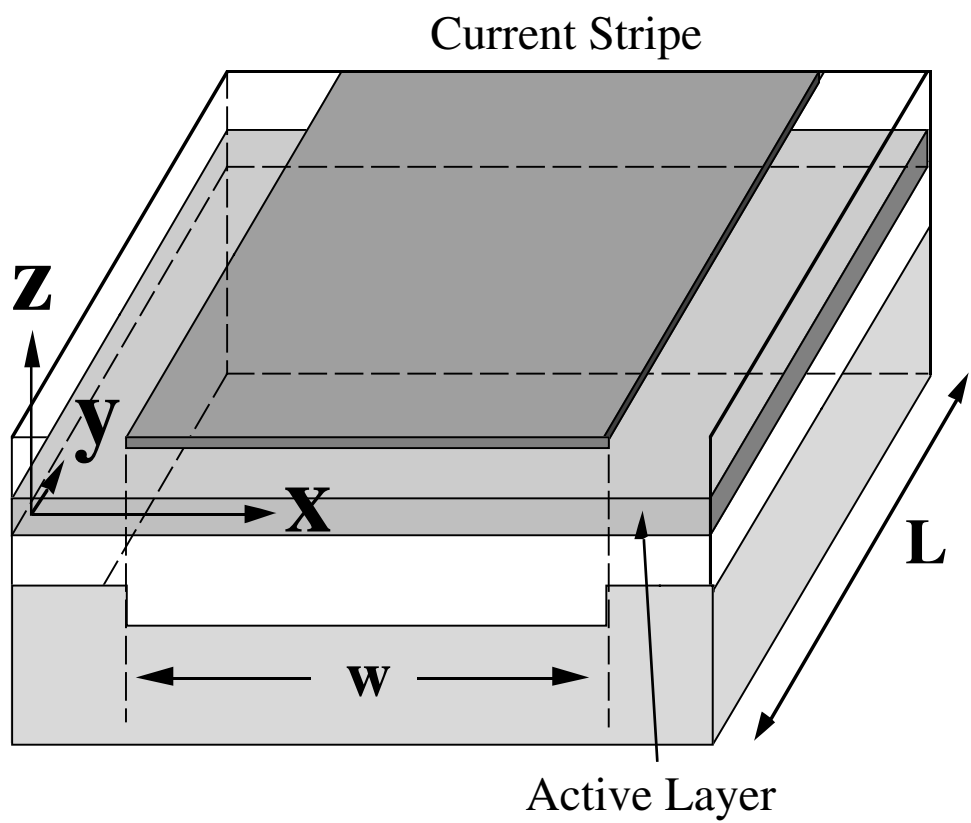
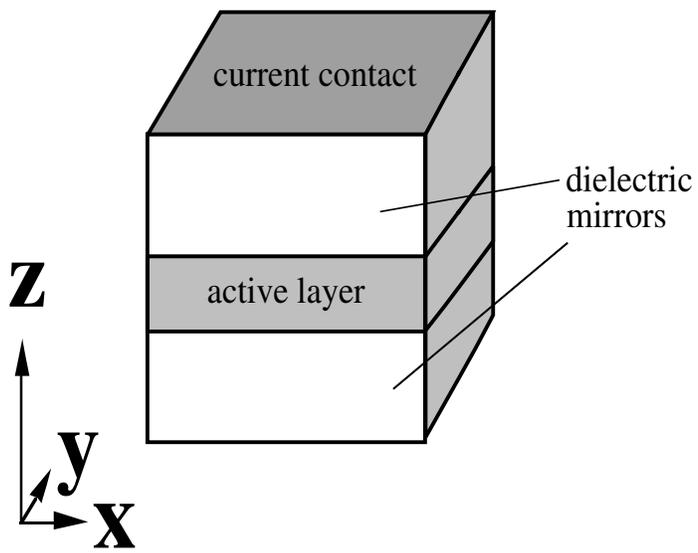
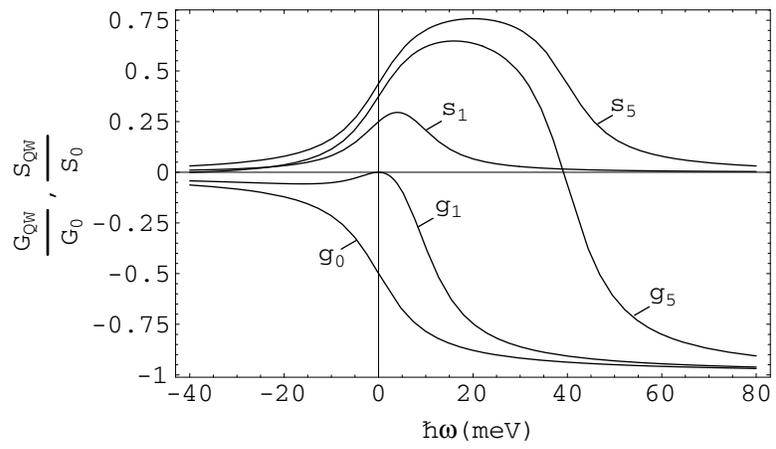
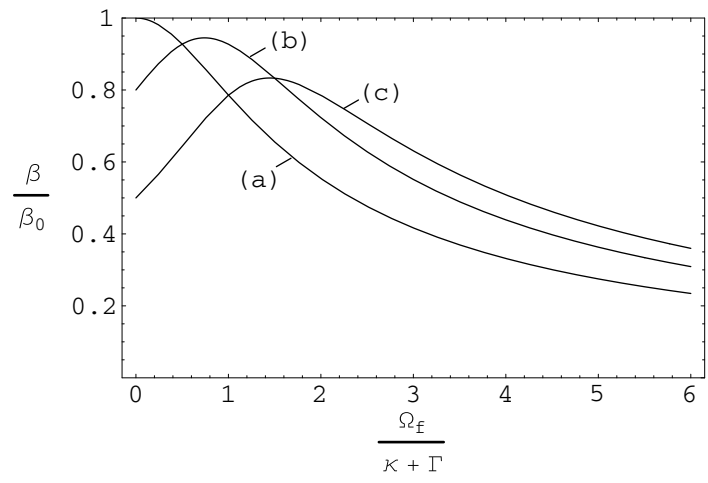
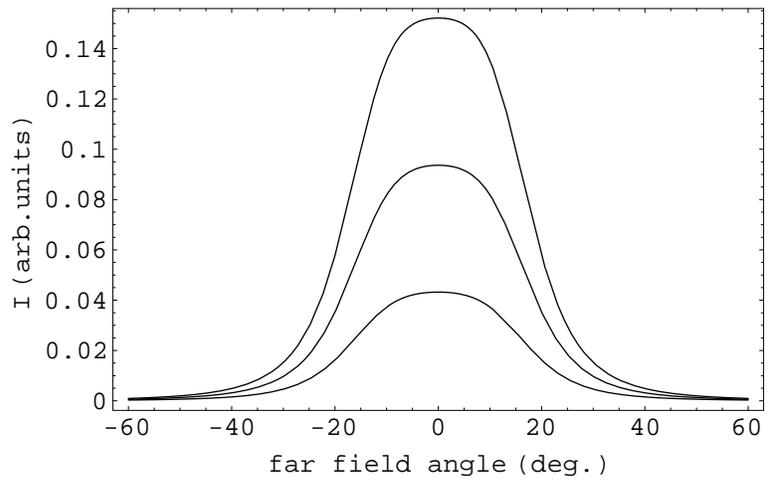
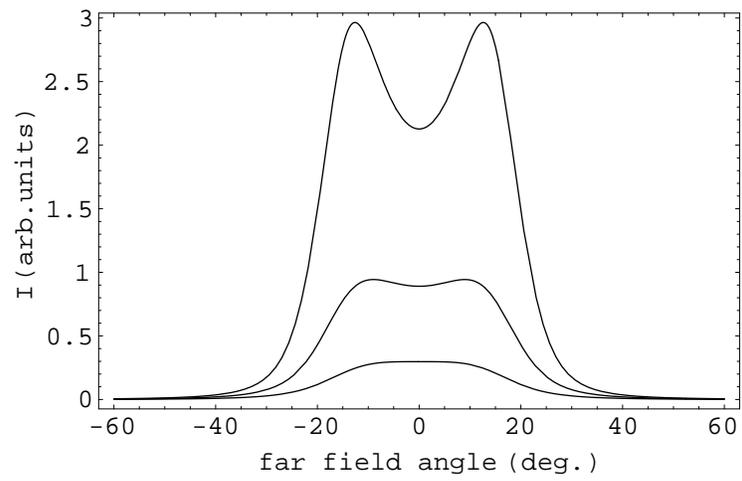
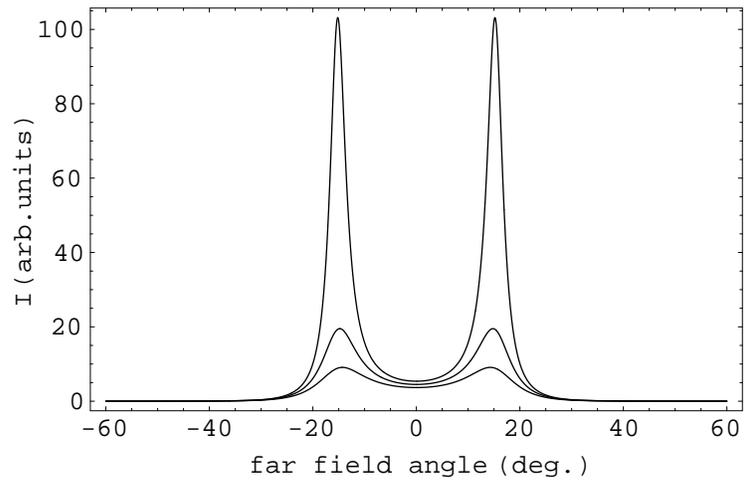
\end{document}